\documentclass{aa}  

\usepackage{graphicx}
\usepackage{txfonts}
\usepackage{bigints}
\usepackage{subfigure}

\begin{document}

   \title{Radiation from the impact of broad-line region clouds onto AGN accretion disks}

   \subtitle{}

   \author{A. L. M\"uller
          \inst{1,3,4}
          \and
          G. E. Romero\inst{1,2} 
          }

   \institute{Instituto Argentino de Radioastronom\'ia (CCT-La Plata, CONICET; CICPBA), 
               C.C. No. 5, 1894, Villa Elisa, Argentina\\
              \email{almuller@iar.unlp.edu.ar}
         \and
             Facultad de Ciencias Astron\'omicas y Geof\'isicas, Universidad Nacional de La Plata, 
             Paseo del Bosque, 1900, La Plata, Argentina\\
             \email{romero@iar.unlp.edu.ar}
         \and
             Institute for Nuclear Physics (IKP), Karlsruhe Institute of Technology (KIT), Germany
         \and
             Instituto de Tecnolog\'ias en Detecci\'on y Astropart\'iculas (CNEA, CONICET, UNSAM), Buenos Aires, Argentina
             }

   \titlerunning{Radiation from BLR clouds onto AGN accretion disks}
   \authorrunning{A. L. M\"uller and G. E. Romero}

   \date{\today}

 
  \abstract
   {Active galactic nuclei are supermassive black holes surrounded by an accretion disk, two populations of clouds, bipolar jets, and a dusty torus. The clouds move in Keplerian orbits at high velocities. In particular, the broad-line region (BLR) clouds have velocities ranging from $1000$ to $10000$ km s$^{-1}$. Given the extreme proximity of these clouds to the supermassive black hole, frequent collisions with the accretion disk should occur.}
   {The impact of BLR clouds onto the accretion disk can produce strong shock waves where particles might be accelerated. The goal of this work is to investigate the production of relativistic particles, and the associated non-thermal radiation in these events. In particular, we apply the model we develop to the Seyfert galaxy \object{NGC 1068}.}
   {We analyze the efficiency of diffusive shock acceleration in the shock of colliding clouds of the BLR with the accretion disk. We calculate the spectral energy distribution of photons generated by the relativistic particles and estimate the number of simultaneous impacts needed to explain the gamma radiation observed by \textit{Fermi} in Seyfert galaxies.}
   {We find that is possible to understand the measured gamma emission in terms of the interaction of clouds with the disk if the hard X-ray emission of the source is at least obscured between $20\%$ and $40\%$. The total number of clouds contained in the BLR region might be between $3\times10^{8}$ and $6\times10^{8}$, which are values in good agreement with the observational evidence. The maximum energy achieved by the protons ($\sim$ PeV) in this context allows the production of neutrinos in the observing range of IceCube.}
   {}

   \keywords{radiation mechanisms: non-thermal --
                shock waves --
                galaxies: actives --
                galaxies: individual (NGC 1068)
               }

\maketitle
%

\section{Introduction}\label{sec:intro}

\begin{figure}
  \centering
  \resizebox{\hsize}{!}{
  \includegraphics[width=0.40\textwidth]{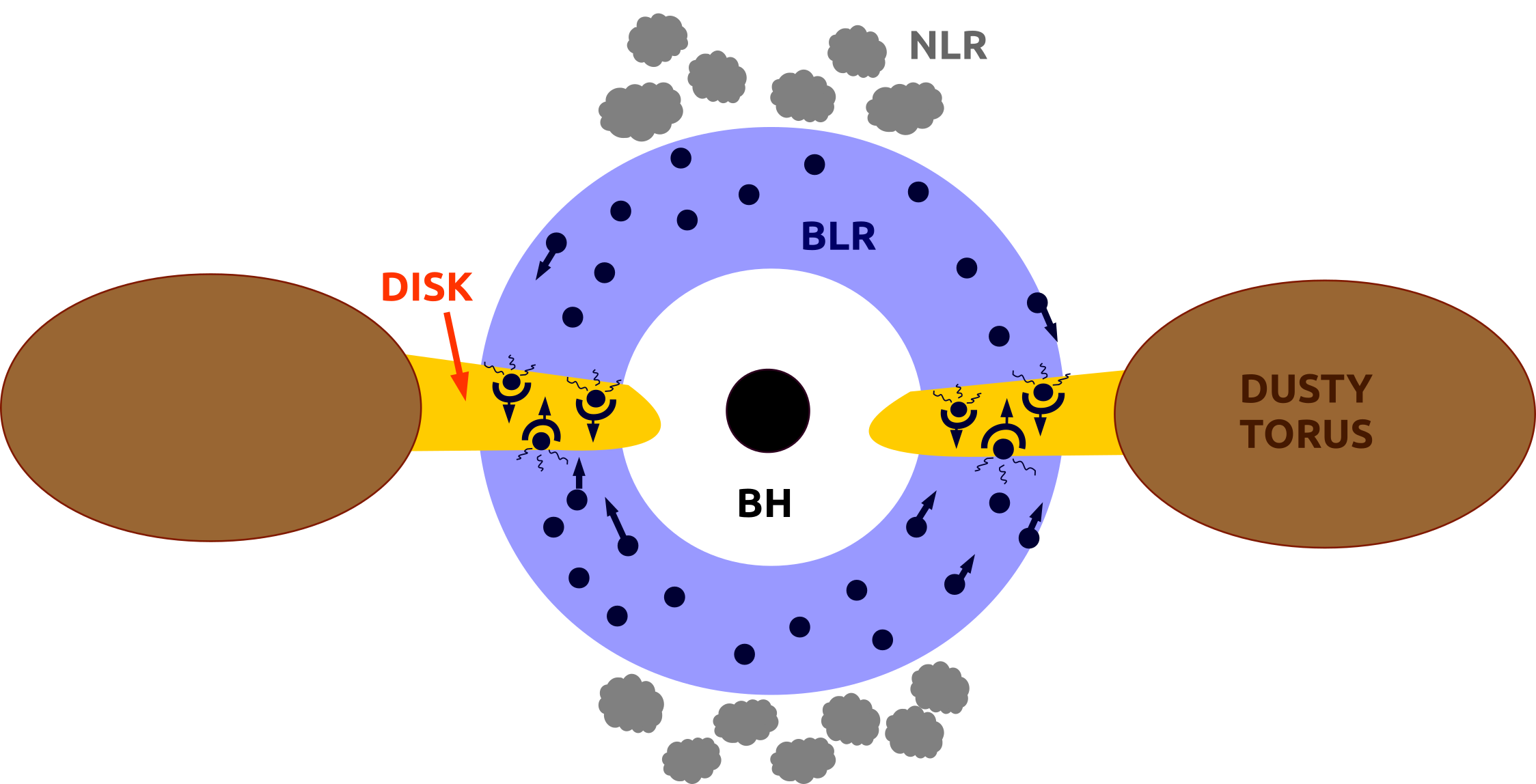}}
  \caption{Illustrative sketch of the physical situation.}
  \label{fig:schemeBLR}
  \end{figure}

Active galactic nuclei (AGNs) are formed by accreting supermassive black holes (BHs). Their characteristic emission is produced by a very compact region and covers a wide range of frequencies \citep{PT2002}. From an observational point of view, objects defined as AGNs are actually very diverse. A first distinction can be made between radio-quiet and radio-loud AGNs. The latter are bright radio objects whose radiation in that band is several orders of magnitude larger than the typical emission of the radio-quiet nuclei. These two main groups can be additionally divided taking into account a variety of characteristics (e.g., the alignment of the jet with the line of sight, the intensity of the lines in the spectra, their luminosity; see \citet{Dermer2016}). The heterogeneity of AGNs can be understood in terms of a unified model adjusting the orientation to the observer and the values of parameters related to the central BH \citep{Antonucci1993,Urry1995}. In unified pictures, an AGN is essentially a supermassive BH surrounded by a subparsec accretion disk and a dusty torus. Inside the torus two populations of clouds move in Keplerian orbits: the broad-line region (BLR) and the narrow-line (NLR) region clouds (see \mbox{Fig. \ref{fig:schemeBLR}}). In the case of radio-loud AGNs, the system also includes  a relativistic jet emitting  synchrotron radiation.

The ultraviolet (UV) and optical spectra of some subclasses of AGNs have prominent broad emission lines (e.g., Seyfert 1). The gas producing these lines should be contained in a central region close to the BH. The structure of this zone is modeled as a group of clouds orbiting in random directions, but with velocities in the range of \mbox{$\sim 10^{3}$ km s$^{-1}$} to \mbox{$\sim 10^{4}$ km s$^{-1}$} \citep{Blandford1990}. The electron number density of the BLR clouds ranges typically from \mbox{$10^{9}$ cm$^{-3}$} to \mbox{$10^{13}$ cm$^{-3}$} and the gas is completely photoionized by the disk radiation. The BLR reprocesses around $10\%$ of the disk luminosity and re-emits lines with a mean energy of \mbox{$10$ eV} and a typical photon density of \mbox{$\sim 10^{9}$ cm$^{-3}$} \citep{Abol2017}. 

The nucleus of Seyfert 2 galaxies is typically obscured by the dusty torus. Therefore, the BLR appears partially hidden but still detectable in the spectropolarimetric data \citep[see, e.g.,][]{Antonucci1984,AntonucciMiller1985,RamosAlmeida2016}. Only in the low-luminosity Seyfert 2 AGNs the existence of the BLR has not been confirmed by observational data \citep{Laor2003,Marinucci2012}.

Since in the standard AGN model the BLR clouds co-exist with the accretion disk, and given the strong evidence of infall motion \citep{Doroshenko2012,Grier2013}, direct collisions between these clouds and the disk should occur. Similarly, the interaction of stars and BHs with the accretion disk has been analyzed before by many authors, but with emphasis on the AGN fuelling consequences, the thermal emission, and the gravitational waves produced in the impacts \citep{Zentsova1983,Syer1991,Colgate1994,Armitage1996,Sillanpaa1988,Nayakshin2004,Doenmez2006,Valtonen2008}. 

In this work, we study the possibility of accelerating particles by first-order Fermi mechanism in the shock waves produced by the impacts of BLR clouds with the accretion disk (Section \ref{sec:basicmodel}). In Sections \ref{sec:partaccenergylosses}, \ref{sec:partdistr}, and \ref{sec:sed}, we present estimates of the cosmic ray acceleration inside the shocked cloud and model the non-thermal emission. Finally, we apply our model to the Seyfert galaxy \object{NGC 1068} (Section \ref{sec:NGC1068}) and discuss the contribution of these impacts to the high-energy radiation in Section \ref{sec:discussion}.

\section{Basic model}\label{sec:basicmodel}

  We assume a standard AGN with a central Schwarzschild BH of mass \mbox{$10^{8}$ M$_{\odot}$} surrounded by a Shakura-Sunyaev accretion disk \citep{SS1973}. The disk extends from the last stable orbit\footnote{$R_{\rm ISCO}=6 R_{\rm g} \sim 3 \times 10^{-5}$ pc} to hundreds of thousands of gravitational radii \citep{APAp}. We adopt standard values for the accretion efficiency and viscosity parameters, namely $\eta_{\rm accre}=0.1$ and $\alpha_{\rm accre} \sim 0.1$, respectively \citep{APAp,Fabian1999,Xie2009}. The bolometric luminosity is $\sim 7\times10^{45}$ erg s$^{-1}$, i.e., $\lambda_{\rm Edd}=L_{\rm bol}/L_{\rm Edd}=0.7$. We calculate the characteristic values of the different parameters of the disk at each radius using the expressions provided by \citet{TMA1988}. The spectrum of the accretion disk is obtained integrating the Planck function over the surface area. Each disk ring has a characteristic temperature. From the resulting expression, the total luminosity $L_{\rm disk}$ is calculated integrating the spectrum over the whole energy range of the emission.
 
  The clouds in the BLR are considered to be spherical and homogeneous with a radius \mbox{$R_{\rm c}=2 \times 10^{13}$ cm} \citep{S2015}. The parameters adopted for an average cloud in our model are shown in Table \ref{tab:cloudparameters}. The evidence obtained from many observational studies indicates that the clouds existing in the BLR move in Keplerian orbits with velocities between $10^{3}$ and \mbox{$10^{4}$ km s$^{-1}$} \citep{Blandford1990,Peterson1998}. We adopt a scenario where the cloud velocity is \mbox{$v_{\rm c}=5000$ km s$^{-1}$}. This speed corresponds to a circular Keplerian orbit of radius \mbox{$r=5.40 \times 10^{16}$ cm $=0.02$ pc}, which is the distance from the galactic center to the place where the impact of the cloud on the equatorial disk occurs. The relevant physical properties of the disk at that radius are shown in Table \ref{tab:discparameters}.

  \begin{table}[htbp]
        \centering
        \caption{Initial parameter values of a BLR cloud.}
        \begin{tabular}{l c}
                \hline\hline
                Parameter [units] & Value \\ \hline
                $R_{\rm c}$ cloud radius [cm] &  $2.0 \times 10^{13}$  \\
                $\rho_{\rm c}$ volumetric density [g cm$^{-3}$] &  $2.2 \times10^{-14}$ \\
                $z$ chemical composition [z$_{\odot}$] &  $1.0$  \\
                $n_{\rm e}$ electron number density [cm$^{-3}$] & $10^{10}$ \\
                $n_{\rm c}$ number density [cm$^{-3}$] &  $1.3 \times10^{10}$  \\
                $M_{\rm c}$ cloud mass [M$_{\odot}$] &  $3.6 \times10^{-07}$ \\
                $v_{\rm c}$ cloud velocity [km s$^{-1}$] & 5000\\ \hline
        \end{tabular}
        \label{tab:cloudparameters}
  \end{table}

  The size of the BLR ranges typically from $0.01$ to \mbox{$1$ pc} \citep{Allen}. One way to estimate this quantity is through reverberation studies \citep[see][]{Kaspi2007}. The values obtained can differ by about an order of magnitude using different emission lines \citep{PetersonW1999}. Therefore, it is necessary to account for a wide range of radii (e.g., from $10^{-3}$ to $0.1$ pc) to reproduce the line pattern attributed to a BLR \citep{Abol2017}. We  assume that the BLR is a thin shell, whose internal radius is $\sim r$, whereas the external one is given by $R_{\rm BLR}=\sqrt{0.1\,L_{\rm disk}/(\pi\,U_{\rm BLR}\,c)}$ \citep{Boettcher2016}. 
  
    \begin{table}[htbp]
        \centering
        \caption{Values of the parameters of the central BH and the associated accretion disk in the model.}
        \begin{tabular}{l c}
                \hline\hline 
                Parameter [units] & Value \\ \hline
                $M_{\rm BH}$ [M$_{\odot}]$ & $10^{8}$ \\
                $\lambda_{\rm Edd}$ Eddington ratio & $0.70$ \\
                $L_{\rm bol}$ bolometric luminosity [erg s$^{-1}$] & $\sim 7\times10^{45}$ \\ \hline
                $r$ impact distance [cm] & $5.40\times10^{16}$ \\ 
                $\eta_{\rm accre}$ accretion efficiency & $0.10$ \\
                $\alpha_{\rm accre}$ viscosity parameter & $0.10$ \\
                $\dot{M}$ accretion rate [M$_{\odot}$ yr$^{-1}$] & $1.54$ \\
                $w_{\rm d}$ disk width [cm] &  $2.44\times10^{14}$ \\
                $\sigma_{\rm d}$ superficial density [g cm$^{-2}$] &  $2.60\times10^{5}$ \\
                $\rho_{\rm d}$ volumetric density [g cm$^{-3}$] &  $1.08\times10^{-9}$  \\
                $n_{\rm d}$ number density \tablefootmark{a} [cm$^{-3}$] &  $6.45\times10^{14}$  \\  
                $L_{\rm disk}$ disk luminosity [erg s$^{-1}$] & $\sim 1.15\times10^{44}$ \\
                $T_{\rm disk}$ temperature [K] &  $1970.70$ \\ \hline
        \end{tabular}
        \tablefoot{
                \tablefoottext{a}{Assuming a disk mainly composed of neutral hydrogen.}}
        \label{tab:discparameters}
  \end{table} 

  The cloud moves supersonically. The collision of the cloud with the accretion disk produces two shock waves: a forward shock propagating through the disk and a reverse shock propagating through the cloud. The velocities of the shocks are calculated with the expressions presented in \citet{LK1996}, whereas the values of the physical parameters in the shocked regions are obtained using the equations for strong adiabatic shocks deduced from the Rankine-Hugoniot relations \citep[see, e.g.,][]{LL1959}. Similar collisions between high-velocity clouds and galactic disks have been studied by several authors \citep[see, e.g.,][]{TT1980,Santillan2004,delValle2018}. 
  
  Adiabatic shocks can be defined demanding that their cooling length $R_{\Lambda}$ is greater than the length of the traversed medium (i.e., the cloud radius and the width of the disk). We calculate the cooling length using the following expression \citep{TT1980}:
  
  \begin{equation}
  \centering
    R_{\Lambda}=\frac{5.06\times10^{-29}(U/\textrm{km s$^{-1}$})^{3}\,A}{(n/ 
\textrm{cm$^{-3}$})\,(\Lambda(T)/\textrm{erg cm$^{3}$ s$^{-1}$})}\textrm{ pc}
  \end{equation}
  with
   \begin{equation}
  T=22\,A\, \left( \frac{U}{\textrm{km s$^{-1}$}} \right)^{2} \text{ K,}
  \label{eqn:longtermica}
  \end{equation}
  
  \noindent where $U$ is the shock velocity with respect to the undisturbed medium of density $n$ and $A$ is a parameter which depends on the conditions of the unshocked gas; its value is $0.6$ if the medium is ionized or $1.6$ if it is neutral. In addition, the function $\Lambda(T)$ is the cooling rate \citep{RCS1976,MZB1998}. 

  The gas in the acceleration region should not be magnetically dominated, otherwise the medium becomes mechanically incompressible and a strong shock cannot exist \citep[see, e.g.,][]{Komissarov2007,Vink2014}. Consequently, the magnetic energy of the medium ($u_{\rm B}$) must be in subequipartition with the kinetic energy of the shocked gas ($u_{\rm g}$), and the magnetization parameter $\beta=u_{\rm B}/u_{\rm g}$ becomes less than $1$. Taking this into account, we assume a modest value of 0.1 for $\beta$ in order to grant effective shock generation and derive the magnetic field in the cloud from

  \begin{equation}
   u_{\rm B}=\beta\,u_{\rm g}
   \nonumber
  \end{equation}

  \begin{equation}
   \frac{B^{2}}{8\pi}=0.1\,\frac{9}{8}n_{\rm c}\,m_{\rm p}\,U^{2} \text{.}
   \label{eq:magnetization10}
  \end{equation}

  \noindent Table \ref{tab:shockedmediums} shows the values of the physical parameters for the shocked media.
  
  \begin{table}[htbp]
  \centering
  \caption{Nature of the shock and parameter values of the adiabatic media.}
  \begin{tabular}{l c c}
  \hline\hline
  Parameter {[}units{]} & Cloud & Disk \\ \hline
   $U$  {[}km s$^{-1}${]} &  $6631$ & $36$ \\
   $R_{\Lambda}$ cooling distance {[}cm{]} & $2.5\times10^{13}$ & $9\times10^{1}$ \\
   Nature of the shock &  adiabatic & radiative \\
   $T$ temperature {[}K{]} &  $6\times10^{8}$ & $-$ \\
   $B$ magnetic field {[}G{]} &  $198$ & $-$ \\
   $n$ number density {[}cm$^{-3}${]} & $5.2\times10^{10}$ & $-$ \\ \hline
  \label{tab:shockedmediums}
  \end{tabular}
  \end{table}  
  
  Since the shock moving through the disk turns out to be radiative, it is not efficient enough to accelerate particles. Therefore, we study the cosmic ray production only in the shock that propagates through the cloud. The collision ends after \mbox{$t_{\rm coll}\sim 3.4 \times 10^{4}$ s}, when the shock finally reaches the total length of the cloud. After this time, hydrodynamic instabilities may become important and destroy the cloud \citep{Araudo2010}. In the case of magnetized clouds, it is possible for them to survive up to $\sim 4\,t_{\rm coll}$ or even longer \citep[see][and references therein]{Shin2008}.

\section{Particle acceleration and energy losses}\label{sec:partaccenergylosses}

   First-order Fermi mechanisms can operate in scenarios with strong adiabatic shock waves \citep{Bell1978,Blandford1978}. The acceleration rate for a particle of energy $E$ and charge $q$ in a region with a magnetic field $B$ and where diffusive shock acceleration (DSA) takes place is 
   
   \begin{equation}
    t_{\rm acc}=E\left( \frac{dE}{dt}\right)^{-1}=\eta^{-1} \frac{E}{q\,c\,B}\text{.}
   \end{equation}

   \noindent Here \mbox{$\eta \leq 1$} is the efficiency of the process. Under conditions of the  first-order Fermi mechanism \citep{Drury1983}
   
   \begin{equation}
    \eta^{-1} \sim 20 \frac{D}{r_{\rm g}\,c} \left( \frac{c}{V_{\rm s}} \right)^{2}\text{,}
   \end{equation}

   \noindent where $D$ is the diffusion coefficient of the medium and \mbox{$r_{\rm g}=E/(q\,B)$} is the gyroradius of the particle. We assume that the diffusion proceeds in the Bohm regime, which means that \mbox{$D_{\rm B}=r_{\rm g}\,c/3$}.

    Given that the acceleration can be suppressed in very high-density media, it is necessary to evaluate the importance of the Coulomb and ionization losses suffered by the particles \citep{Drury1996}. In order to evaluate this, we calculate the corresponding cooling times using the expressions provided by \citet{Schlickeiser2002}

      \begin{equation}
          t^{e}_{\rm ion}=1.3\times10^{8}\,\left(\frac{n_{\rm e}}{{\rm cm}^{-3}}\right)^{-1}\,\left(\frac{E}{\rm eV}\right)\,\left[ \ln{\left(\frac{E/{\rm eV}}{n_{e}/{\rm cm}^{-3}}\right)}+61.15 \right]^{-1}\,\textrm{s}
      \end{equation}
     
      \begin{equation}
      \begin{split}
          t^{p}_{\rm ion}=& 3.2\times10^{6}\,Z^{-2}\,\left(\frac{n_{\rm e}}{{\rm cm}^{-3}}\right)^{-1}\,\left(\frac{E}{\rm eV}\right)\\
          &\times\left(\frac{\beta^{2}}{2.34\times10^{-5}\,x_{m}^{3}+\beta^{3}}\right)^{-1}\,\Theta(\beta-7.4\times10^{-4}\,x_{m})\,\textrm{s},
      \end{split}
      \end{equation}
      \noindent where $\Theta$ is the Heaviside function,  $\beta=\sqrt{1-{\gamma^{-2}}}$ (with $\gamma$ the Lorentz factor of the particle), and $x_{m}=(T_{e}/2\times10^{6}\,{\rm K})^{1/2}$.

    The relativistic particles injected lose energy due to the interaction with the matter, photon, and magnetic fields of the cloud. We consider the synchrotron losses (sync) produced by the interaction of the electrons with the magnetic field and the relativistic Bremsstrahlung losses (BS) produced by the interaction of the same particles with the ionized hot matter of the cloud. We also calculate the inverse Compton (IC) upscattering of the photons from the BLR, the accretion disk, and the synchrotron radiation (SSC). The local emission from the disk is approximately a blackbody, whose temperature is $T_{\rm disk}=1970.7$ K. On the other hand, the BLR radiation is a monochromatic photon field with \mbox{$<\epsilon>=10$ eV} and \mbox{$n_{\rm ph} \sim 6.24\times10^{19}$ erg$^{-1}$ cm$^{-3}$}. For the protons, the most relevant radiative process is the proton-proton inelastic collisions ($pp$). We calculate the cooling timescales associated with these processes, using the expressions presented by \citet{Romero2010} (see Eqs. 5-12). 
     
      We also take into account the fact that the particles can escape from the region of acceleration because of diffusion. The cooling rate for this process is \citep{Romero2011}
      
      \begin{equation}
      t_{\rm diff}=\frac{X^{2}}{D}\text{,}
      \end{equation}
      
      \noindent where $D$ is the diffusion coefficient of the medium (i.e., the Bohm diffusion coefficient in our model) and $X$ is the characteristic size of the acceleration region. We assume $X=R_{\rm c}$.
  
      Another non-radiative process that we include is the adiabatic loss, i.e., the energy loss due to the work done by the particles expanding the shocked cloud matter. The cooling timescale is given by \citep{Bosch-Ramon2010}
      \begin{equation}
      t_{\rm adi}=4.9 \times 10^{-8} \left(\frac{U}{10^{3}\,\textrm{km s}^{-1}}\right)^{-1}\,\left(\frac{R_{\rm c}}{\rm cm}\right)\,\textrm{s.}
      \end{equation}
      
      The maximum energy that the particles can reach before they escape from the acceleration region is constrained by the Hillas criterion \mbox{$E_{\rm max}=X\,q\,B$} \citep{Hillas1984}. The maximum value according to this criterion is \mbox{$\sim 3\times 10^{18}$ eV}.
      
\begin{figure*}
       \centering
       \subfigure[electrons]{\includegraphics[trim={1.7cm 1.8cm 1cm 1.8cm},clip,width=0.49\textwidth]{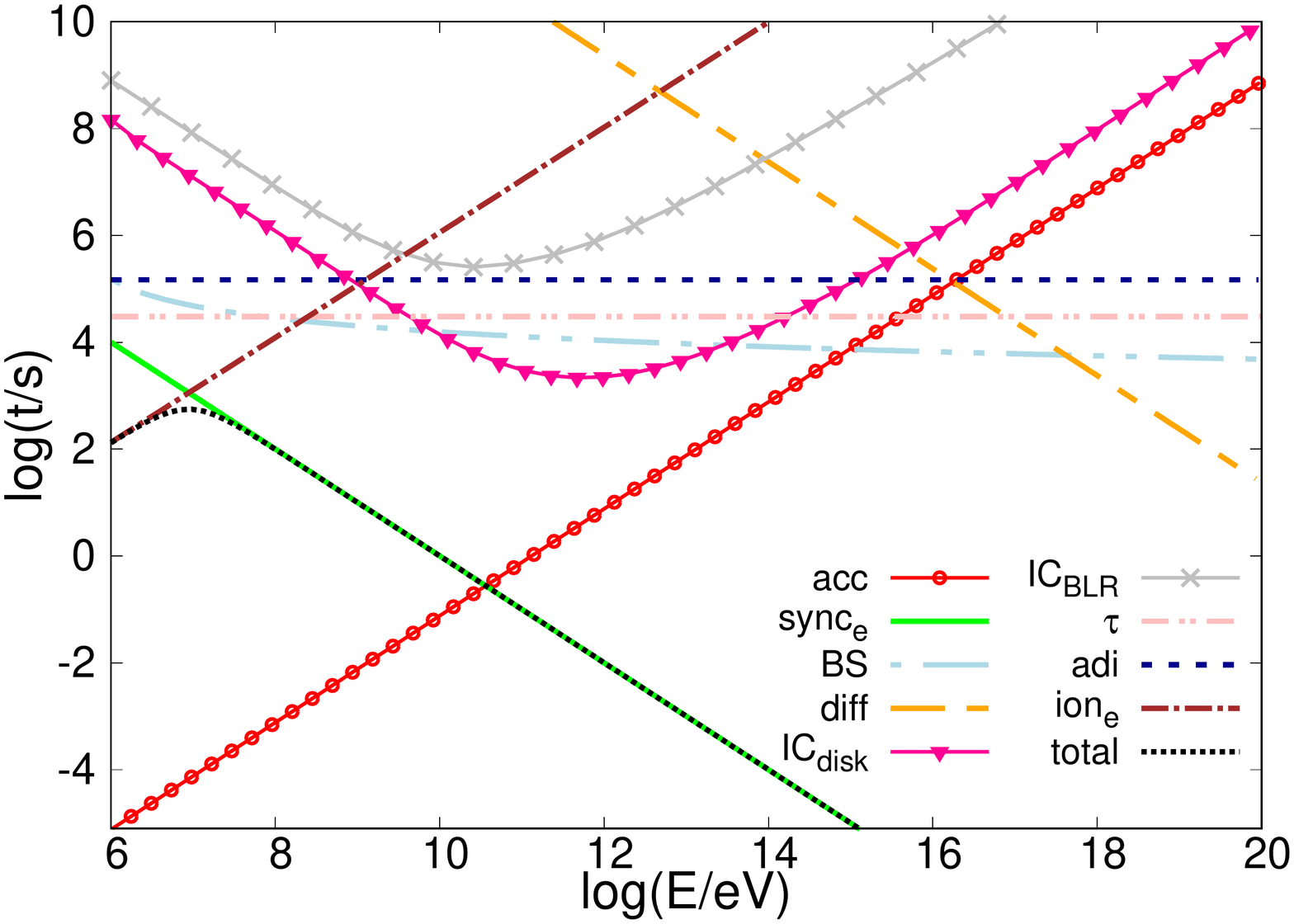}\label{fig:coolingratee}}
       \subfigure[protons]{\includegraphics[trim={1.7cm 1.8cm 1cm 1.8cm},clip,width=0.49\textwidth]{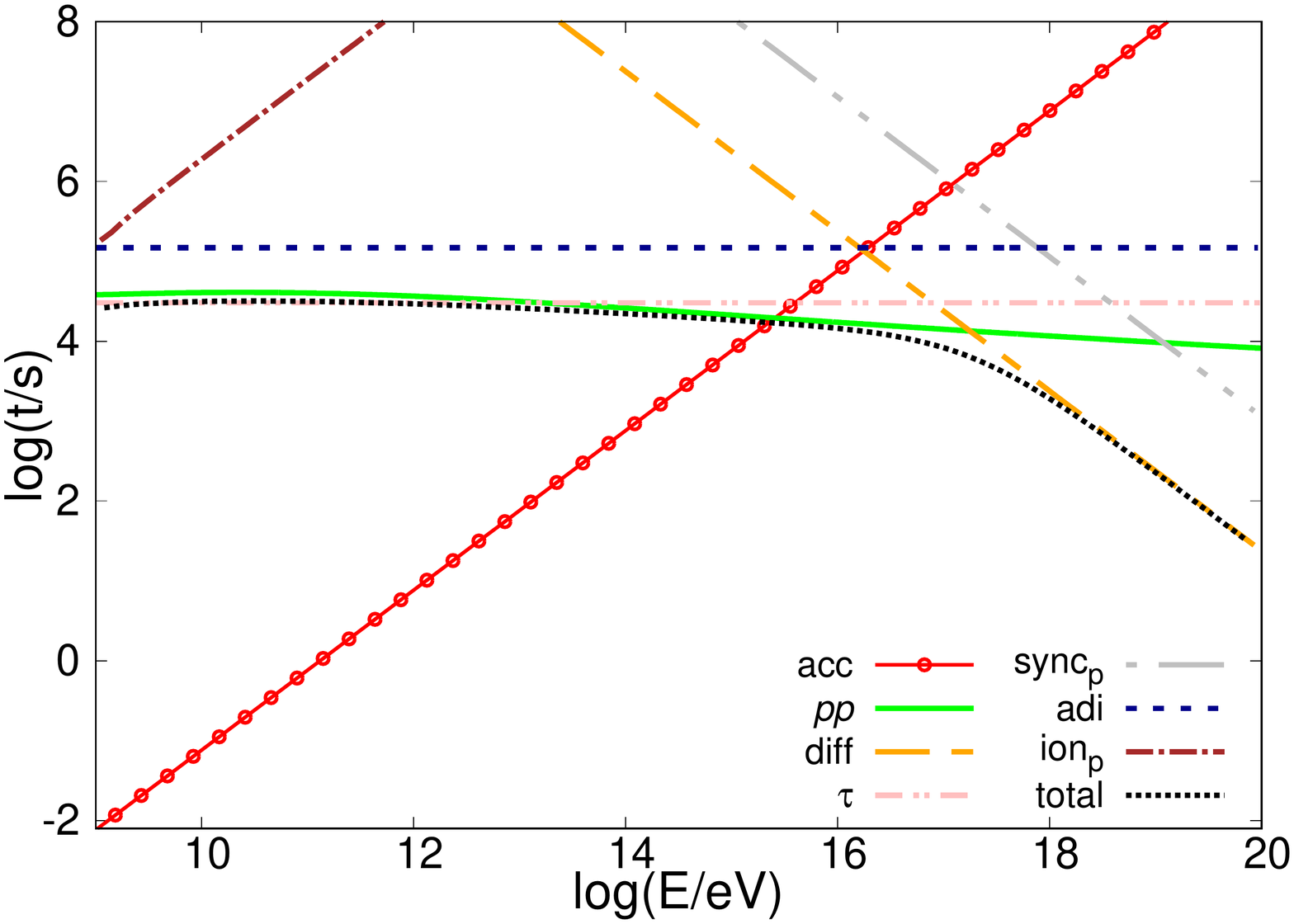}\label{fig:coolingratep}}
       \caption{Cooling and acceleration timescales for the particles, where $\tau=X/U$ is the age of the source.}
       \label{fig:coolingrates}
       \end{figure*}

      In Fig. \ref{fig:coolingratee} and Fig. \ref{fig:coolingratep} we show the cooling timescales together with the diffusion and acceleration rates for the electrons and protons. The maximum energy for each kind of particle can be inferred looking at the point where the acceleration rate is equal to the total cooling and/or escape rate. 
      
      Synchrotron dominates the energy losses for the electrons over the whole energy range, whereas the IC losses become negligible (see Fig. \ref{fig:coolingratee}). This is expected because the magnetic energy density in the cloud \mbox{U$_{\rm mag}=1.54\times10^{3}$ erg cm$^{-3}$} is much higher than the blackbody radiation energy density \mbox{U$_{\rm disk}=1.31\times10^{-1}$ erg cm$^{-3}$} of the disk and the photon density of the BLR \mbox{U$_{\rm BLR}=1.60\times10^{-2}$ erg cm$^{-3}$}. For protons, the $pp$ dominates the energy losses (see Fig. \ref{fig:coolingratep}). Consequently, the maximum energies are \mbox{$E^{e}_{\rm max}=3.6\times 10^{10}$ eV} and \mbox{$E^{p}_{\rm max}=1.5\times10^{15}$ eV} for electrons and protons, respectively. The Hillas criterion is satisfied by particles of such energies, and thus   are the maximum energies that the particles can reach in our scenario. 
      
      Another important result shown by these plots is that, after the end of the collision, the produced cosmic rays  cool down locally and do not propagate. Electrons  lose all their energy almost immediately (\mbox{$\sim 3\times10^{2}$ s}), whereas the protons will lose it after $\sim 3\times10^{4}$ s. This timescale is comparable to $t_{\rm coll}$, thus the accelerated hadrons, and the secondary particles created by them, will emit longer than the primary leptons. 
      
\section{Particle distributions}\label{sec:partdistr}

We solve the transport equation for relativistic particles,

\begin{equation}
  \frac{\partial N_{e,p}(E, t)}{\partial t}+\frac{\partial[b(E) N_{e,p}(E,t)]}{\partial E}+ \frac{N_{e,p}(E, t)}{t_{\textrm{esc}}}=Q_{e,p}(E) \label{Teq}
,\end{equation}

\noindent to find the particle distributions \citep{Ginzburg1964}. In this expression, $Q_{e,p}(E)$ represents the injection term, \mbox{$b(E)=dE/dt$} the sum of all the energy losses, and $t_{\textrm{esc}}$ the escape time of the particles (i.e., the diffusion). For the injection function, we assume a power law with an exponential cutoff $Q_{e,p}(E)=K_{e,p}\,E^{-2}\,\exp{({-E/E^{e/p}_{\rm max})}}$, which corresponds to the injection produced by a DSA in adiabatic strong shocks. Given that the particle distribution of electrons reaches the steady state at once and the proton distribution does the same in only $\sim 10^{4}$ s, we study the steady state solution of the transport equation. The time interval where this solution is valid is $t_{\rm ss}=t_{\rm coll}-10^{4}\textrm{ s}=2.4\times10^{4}$ s.

  The power injected per impact can be calculated as \mbox{$L_{\rm s}=\frac{1}{2}{M_{\rm c} v_{\rm c}^{2}}/{t_{\rm coll}}$} \citep{delValle2018}. The kinetic energy obtained using the set of parameters of our model is \mbox{$L_{\rm s}=3.9 \times 10^{40}$ erg s$^{-1}$}. We assume that 10\% of the energy available is used to accelerate particles up to relativistic energies. Therefore, the power available to accelerate electrons and protons in the cloud is \mbox{$\sim 3.9 \times 10^{39}$ erg s$^{-1}$}. How this luminosity is divided between the electrons $(L_{e})$ and protons $(L_{p})$ is uncertain. We consider two situations: energy equally distributed between the two particle types \mbox{$(L_{p}/L_{e}=1)$} and $100$ times the energy injected in electrons to protons \mbox{$(L_{p}/L_{e}=100)$}.
      
\section{Spectral energy distributions}\label{sec:sed}

Using the particle energy distributions obtained in the previous section (Section \ref{sec:partdistr}), we calculate the spectral energy distribution (SED) taking into account all the radiative processes mentioned and correcting the result by absorption. To this end, we suppose that the emission region is a spherical cap with height $X$, so its volume is $V_{\rm c}={\pi}\,X^{2}\,(3\,R_{\rm c}-X)/3$.

\subsection{Radiative processes}

In the case of the synchrotron emission, we use the expressions provided by \citet{Blumenthal1970}. Then the synchrotron luminosity emitted by a distribution of electrons $N_{\rm e}(E_{\rm e})$ can be calculated as

\begin{equation}
 L_{\rm sync}(E_{\gamma})=E_{\gamma}\,V_{\rm c}\,\kappa_{\rm SSA}(E_{\gamma})\int_{E^{e}_{\rm min}}^{E^{e}_{\rm max}} N_{\rm e}(E)\,P_{\rm sync}(E,E_{\gamma})\,dE\text{,}
\end{equation}

\noindent with

\begin{equation}
P_{\rm sync}(E_{\rm e}, E_{\gamma})=\frac{\sqrt{3}\,e^{3}\,B}{h\,m_{\rm e}\,c^{2}}\,\frac{E_{\gamma}}{E_{\rm c}}\int_{E_{\gamma}/E_{\rm c}}^{\infty} K_{5/3}(\zeta)\,d\zeta
\end{equation}

\noindent and

\begin{equation}
E_{\rm c}=\frac{3}{4\,\pi}\frac{e\,h\,B}{m_{e}\,c}\left(\frac{E}{m_{e}\,c^{2}}\right)^{2}.
\end{equation}

\noindent Here, $K_{5/3}(\zeta)$ is a modified Bessel function. Defining \mbox{$\varphi=E_{\gamma}/{E_{\rm c}}$}, we use that $\varphi\,\int_{\varphi}^{\infty}K_{5/3}(\zeta)\,d\zeta\approx1.85\,\varphi^{1/3}e^{-\varphi}$ if $0.1\le \varphi \le 10$ \citep{Romero2011}. The coefficient $\kappa_{\rm SSA}(E_{\gamma})$ is the correction due to synchrotron self-absorption (SSA):

\begin{equation}
\kappa_{\rm SSA}(E_{\gamma})=\frac{1-e^{-\tau_{\rm SSA}(E_{\gamma})}}{\tau_{\rm SSA}(E_{\gamma})}.
\end{equation}

\noindent The expression for the optical depth $\tau_{\rm SSA}$ can be found in \citet{Rybicki}. 

We calculate the IC emission and Bremsstrahlung\footnote{There is a typo in Eq. 31. The $4\pi$ in the denominator should be removed.} using the expressions presented by \citet{Romero2010b} (Eqs. 28-33). To estimate the gamma luminosity generated by $pp$ inelastic collisions, we follow the procedure given by \citet{Kelner2006} (see Section IV and V). Following this approach, the emissivity produced by protons with $E_{p} < 100$ GeV is obtained using the $\delta$-functional approximation \citep{Aharonian2000}, whereas for $E_{p} > 100$ GeV corrections accounting for the production of charge particles are introduced. Finally, we also calculate and include the thermal contribution from the accretion disk.

\subsection{Absorption and secondary particles}

The interaction of the gamma photons generated by $pp$ collisions with the UV photons from the BLR, and with the optical photons coming from the accretion disk, inject secondary electron-positron pairs.

The optical depth for gamma rays propagating in this scenario can be calculated with the expression for the total cross section provided by \citet{Gould1967}, being the threshold condition for pair production \mbox{$E_{\gamma}\,\epsilon > (m_{\rm e}\,c^{2})^{2}$}. Since we assume $\epsilon=10\,$eV for the BLR photons, gamma rays with $E_{\gamma}>30\,$GeV satisfy this condition. In the case of the absorption by accretion disk photons, the threshold is exceeded by gamma photons with $E_{\gamma}>1.5\,$TeV.

The injection of secondary particles $Q^{\rm sec}_{\rm e}(E_{\rm e})$ (in units of erg$^{-1}$ s$^{-1}$ cm$^{-3}$) produced in photon-photon interactions, if \mbox{$\epsilon \ll m_{\rm e}\,c^{2} \leq E_{\gamma}$}, can be approximated as \citep[see, e.g.,][and the references therein]{Romero2010}

 \begin{equation}
 \begin{split}
 Q^{\rm sec}_{\rm e}(E_{\rm e})=&\frac{3}{32}\frac{c\,\sigma_{\rm T}}{m_{\rm e}\,c^{2}}\bigintsss_{\gamma_{\rm e}}^{\infty}\bigintsss_{\frac{\epsilon_{\gamma}}{4\,\gamma_{\rm e}\,(\epsilon_{\gamma}-\gamma_{\rm e})}}^{\infty} d\epsilon_{\gamma}\,d\omega\,\frac{n_{\gamma}(\epsilon_{\gamma})}{\epsilon_{\gamma}^{3}}\,\frac{n_{\rm ph}(\omega)}{\omega^{2}} \\ 
 & \times \left\{ \frac{4\,\epsilon_{\gamma}^{2}}{\gamma_{\rm e}\,(\epsilon_{\gamma}-\gamma_{\rm e})} \ln \left[ \frac{4\,\gamma_{\rm e}\,\omega\,(\epsilon_{\gamma}-\gamma_{\rm e})}{\epsilon_{\gamma}} \right]-8\,\epsilon_{\gamma}\,\omega \right. \\ 
 & \left. + \frac{2\,(2\,\epsilon_{\gamma}\,\omega-1)\,\epsilon_{\gamma}^{2}}{\gamma_{\rm e}\,(\epsilon_{\gamma}-\gamma_{\rm e})}-\left( 1-\frac{1}{\epsilon_{\gamma}\,\omega}\right) \frac{\epsilon_{\gamma}^{4}}{\gamma_{\rm e}^{2}\,(\epsilon_{\gamma}-\gamma_{\rm e})^{2}} \right\}\text{,}
 \end{split}
 \end{equation} 

 \noindent where $\gamma_{\rm e}=E_{\rm e}/(m_{\rm e}\,c^{2})$, $\epsilon_{\gamma}=E_{\gamma}/(m_{\rm e}\,c^{2})$, and $\omega=\epsilon/(m_{\rm e}\,c^{2})$. These particles interact and emit by the same processes as the primary electrons. Considering that the synchrotron radiation dominates the cooling of the electrons, we only calculate this emission for the secondaries.

\subsection{Results}

\begin{figure*}
       \centering
        \subfigure[$L_{p}/L_{e}=1$]{\includegraphics[trim={1cm 1cm 1cm 1cm},clip,width=0.49\textwidth]{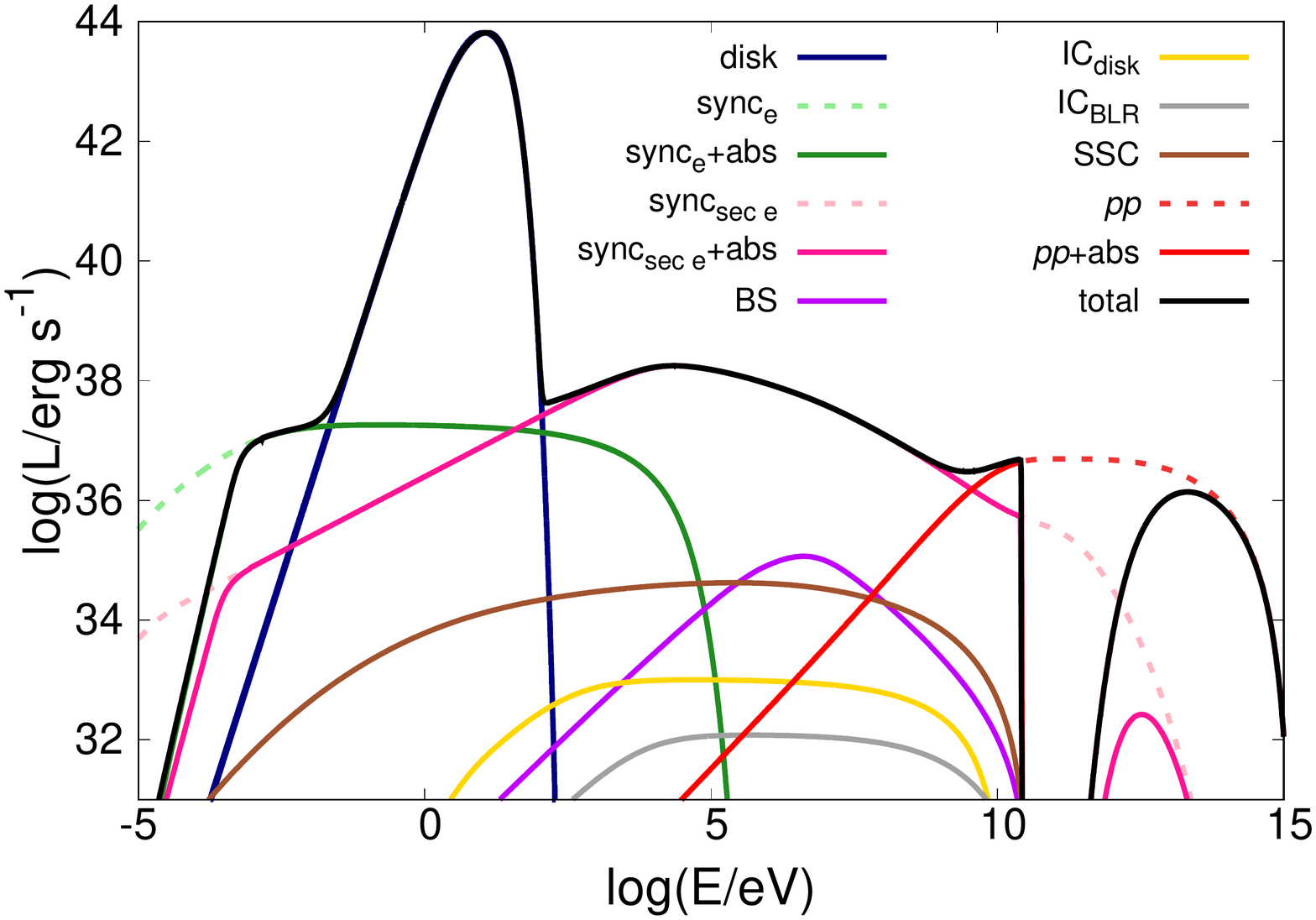}\label{fig:SED_a1model}}
       \subfigure[$L_{p}/L_{e}=100$]{\includegraphics[trim={1cm 1cm 1cm 1cm},clip,width=0.49\textwidth]{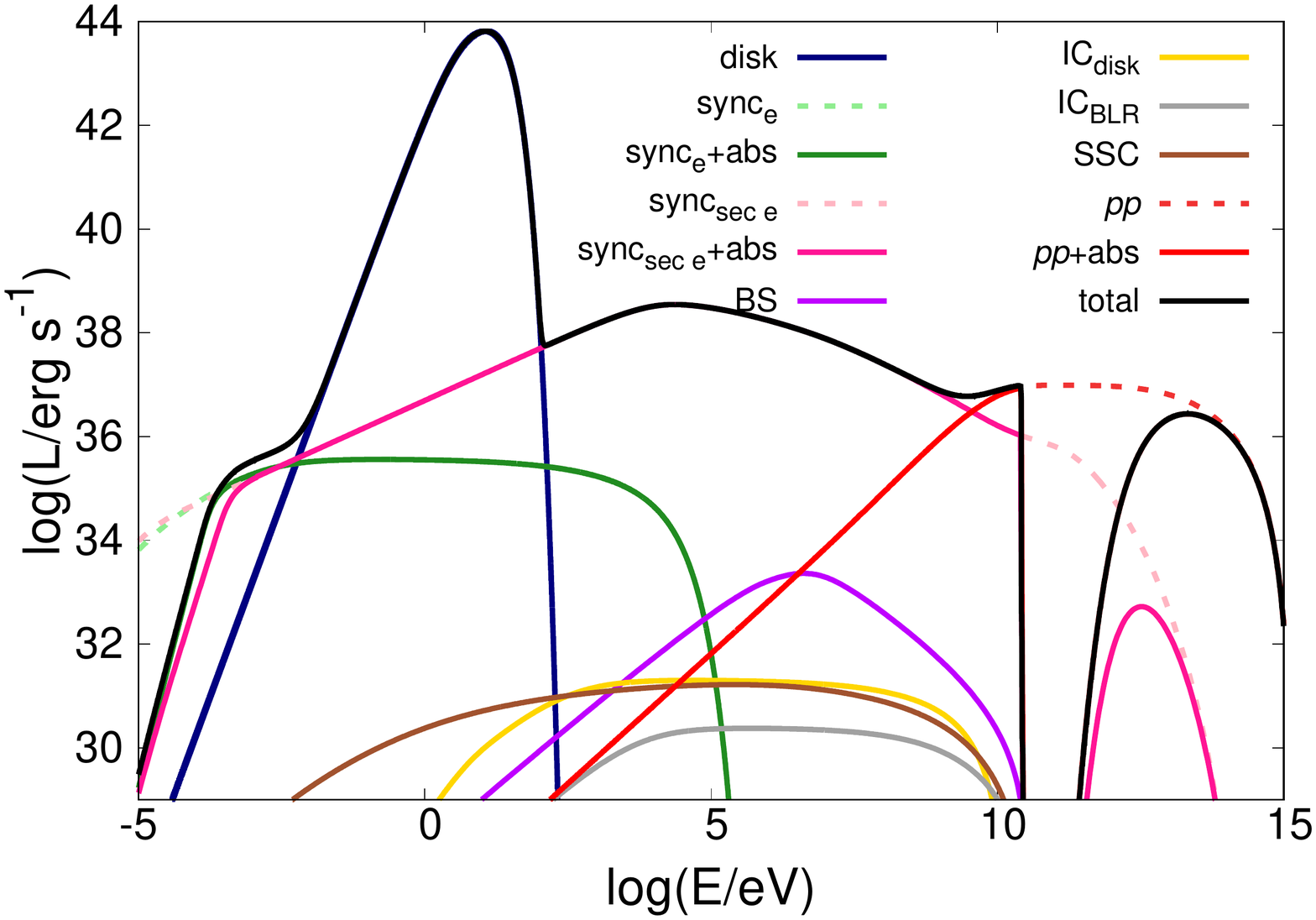}\label{fig:SED_a100model}}
       \caption{Spectral energy distributions obtained for our model. The plot on the left represents the SED obtained with equal power injected in protons than electrons. The plot on the right shows the situation where the luminosity injected in protons is 100 times the luminosity that goes to electrons. The dark-blue line labeled ``disk'' is the thermal emission from the accretion disk.}
       \label{fig:SEDsmodel}
       \end{figure*}

Figures \ref{fig:SED_a1model} and \ref{fig:SED_a100model} show the SEDs obtained for $L_{p}/L_{e}=1$ and $L_{p}/L_{e}=100$, respectively. We find that the luminosity at the lowest frequencies (radio) is particularly sensitive to this ratio. The radiation from primary leptons dominates in this part of the spectrum only if the power that goes to the protons is significantly less than $100\,L_{e}$. The radio luminosities in the cases of $L_{p}/L_{e}=1$ and $L_{p}/L_{e}=100$ differ by about a factor 10.

The optical region of the spectrum is dominated by the thermal radiation from the accretion disk, whereas the high-energy part is non-thermal emission produced as a consequence of the acceleration of hadrons. Most of the gamma emission generated in $pp$ collisions is absorbed and converted to secondary particles. The synchrotron radiation of these secondaries prevails in the energy range from \mbox{$1$ keV} to \mbox{$10$ GeV}, having a maximum of \mbox{$\sim10^{38}$ erg s$^{-1}$} at around \mbox{$10$ keV}.

The hard X-rays and gamma luminosity produced by the collision of one BLR cloud are several orders of magnitude below the values typically detected in AGNs by \textit{Swift}, \textit{INTEGRAL}, and \textit{Fermi}. Therefore, a single  event is not expected to be observed as a flare. The very high-energy gamma-ray tail of a single impact might be detected in the future in nearby sources by the forthcoming Cherenkov Telescope Array (CTA). However, since these photons can be easily absorbed if they travel through a dense visible or IR photon field (e.g., from a stellar association or the emission of the dusty torus), they might be strongly attenuated. Nevertheless, we note that the slope of the SED agrees very well with the observational data of a few galaxies like \object{NGC 1068}, \object{NGC 4945}, and \object{Circinus} \citep{Ackermann2012,Woja2015}. Given that the total number of BLR clouds may be around $10^{8}$ or more, it seems more realistic to think about multiple simultaneous collisions, in which case the observed luminosity will be the sum of the individual events. For this reason, in the next section we apply our model to \object{NGC 1068} and discuss the possibility of simultaneous impacts.

\section{Application to \object{NGC 1068}}\label{sec:NGC1068}

\object{NGC 1068} is a spiral edge-on galaxy in the constellation Cetus whose distance to Earth is $D\sim14.4$ Mpc \citep{Tully1988}. This object is classified as a Seyfert 2 galaxy and inspired the AGN unified model \citep{Antonucci1993}. Its bolometric luminosity is estimated to be \mbox{$\sim 8\times10^{44}$ erg s$^{-1}$} \citep{Pier1994}. Although it is considered a star-forming galaxy, its emission can not be completely explained using a starburst model only \citep{Lamastra2016}.

We apply our model to \object{NGC 1068}, using the parameters provided by \citet{LodatoBertin2003} (see Table \ref{tab:NGC1068parameters}). For the BLR cloud, we assume the same parameters adopted previously. We suppose that the total luminosity will be the radiation produced by a single event multiplied by a number $N_{\rm events}$, which is the number of simultaneous events. We fix $N_{\rm events}$ requiring to match the total gamma emission observed by \textit{Fermi} in the range from $100$ MeV to $100$ GeV, which is \mbox{$L_{0.1-100\,{\rm GeV}}=1.85\pm0.14\times10^{41}$ erg s$^{-1}$} \citep{Fermi2019}. We discuss the validity of this assumption in Section \ref{sec:discussion}.

We compare the multiple-event SED with the radio observations taken with VLBA \citep{Gallimore2004} and ALMA \citep{Garcia2016,Impellizzeri2019}, and the gamma-ray spectra produced with the last \textit{Fermi} catalog (8 yr) \citep{Fermi2019} (see Table \ref{tab:NGC1068obs}). The radio observations are considered upper limits because the fluxes reported correspond to regions whose sizes are far larger than the region we are modeling. Furthermore, the data at 256 GHz is the integrated luminosity in a region of $9.1$ pc \citep{Impellizzeri2019}. On the other hand, the spatial resolution of the observation at $694$ GHz is $4$ pc and consequently the thermal emission from the dusty torus is also included \citep{Garcia2016}. 

\citet{Bauer2015}, based on \textit{NuSTAR} observations, suggested that even the hard X-ray luminosity of \object{NGC 1068} is obscured because of its Compton-thick nature. This scenario was recently reviewed and confirmed by \citet{Zaino2020}. This implies that the measured X-ray emission is not intrinsic, but transmitted by reflections. In this situation, the intrinsic radiation in the source is higher than observed.

\begin{table}[htbp]
        \centering
        \caption{Parameter values for the BH, accretion disk of \object{NGC 1068}, and shock properties.}
        \begin{tabular}{l c}
                \hline\hline 
                Parameter [units] & Value \\ \hline
                $M_{\rm BH}$ [M$_{\odot}]$ & $8\times10^{6}$ \\
                $\lambda_{\rm Edd}$ Eddington ratio & $0.77$ \\
                $L_{\rm bol}$ bolometric luminosity [erg s$^{-1}$] & $8\times10^{45}$ \\
                $\eta_{\rm accre}$ accretion efficiency & $0.06$ \\
                $\alpha_{\rm accre}$ viscosity parameter & $10^{-2}$ \\
                $\dot{M}$ accretion rate [M$_{\odot}$ yr$^{-1}$] & $0.235$ \\ \hline
                $n_{\rm d}$ number density [cm$^{-3}$] &  $2.64\times10^{16}$  \\  
                $L_{\rm disk}$ disk luminosity [erg s$^{-1}$] & $1.70\times10^{43}$ \\
                $T_{\rm disk}$ temperature [K] &  $4311.82$ \\ \hline
                $R_{\rm BLR}$ characteristic BLR radius [cm] & $3.36\times10^{16}$ \\
                $U_{\rm c}$ velocity of the shock [km s$^{-1}$] & $6661$ \\
                $t_{\rm coll}$ collision timescale [s] & $3.43\times10^{4}$   \\
                $t_{\rm ss}$ steady state timescale [s] & $2.43\times10^{4}$    \\
                $r$ impact distance [cm] &  $4.26\times10^{15}$ \\ \hline
        \end{tabular}
        \label{tab:NGC1068parameters}
\end{table} 

\begin{table}[htbp]
        \centering
        \caption{Observational data of \object{NGC 1068}.}
        \begin{tabular}{l c c}
                \hline\hline 
                Freq./Energy & Luminosity & Instrument \\ \hline
                $5$ GHz & $(7.44\pm0.62)\times10^{36}$ erg s$^{-1}$ & VLBA\\
                $8.4$ GHz & $(1.15\pm0.10)\times10^{37}$ erg s$^{-1}$ & VLBA\\
                $256$ GHz & $(8.07\pm0.63)\times10^{38}$ erg s$^{-1}$ & ALMA\\
                $694$ GHz & $(2.38\pm0.17)\times10^{39}$ erg s$^{-1}$ & ALMA \\\hline
                $14-195$ keV & $9.40^{+0.62}_{-0.53}\times10^{41}$ erg s$^{-1}$ & \textit{Swift} \\ \hline
                $0.1-100$ GeV & $(1.85\pm{0.14})\times10^{41}$ erg s$^{-1}$ & \textit{Fermi} 8 yr \\ \hline
        \end{tabular}
        \label{tab:NGC1068obs}
\end{table} 

\begin{figure}
        \centering
        \includegraphics[trim={1.3cm 1.8cm 1.5cm 2.2cm},clip,width=0.49\textwidth]{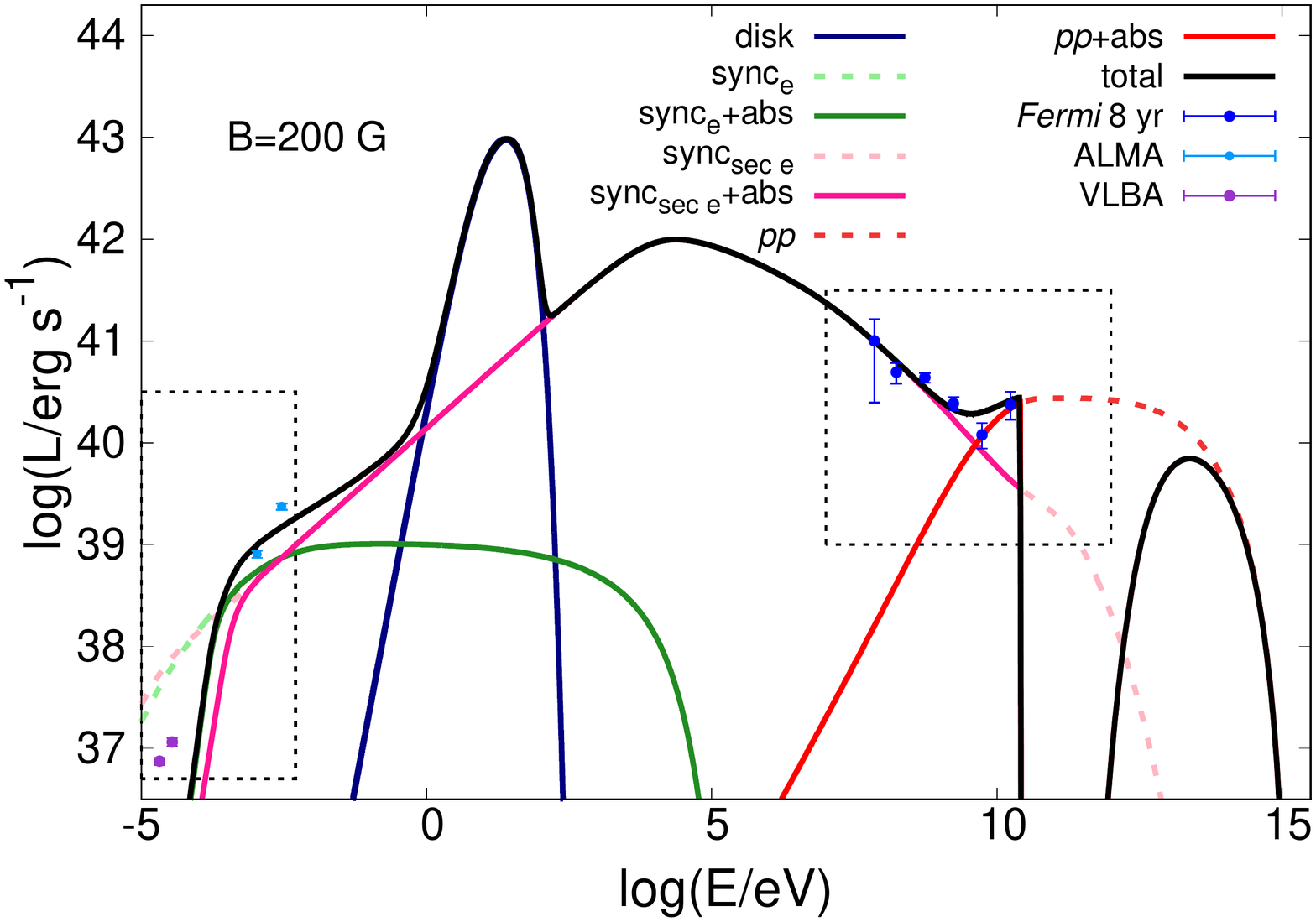}
        \caption{SED assuming $B=200$ G and $L_{p}/L_{e}=100$. The number of cloud impacts, $N_{\rm events}$, is determined by adjusting the model to match the observed total gamma luminosity. The gamma absorption is produced by the UV photons of the BLR and the optical photons from the accretion disk. The regions contained in the dashed rectangles are expanded in Fig. \ref{fig:NGC1068_radio} and Fig. \ref{fig:NGC1068_gamma}.}
        \label{fig:NGC1068_200}
\end{figure}

\begin{figure}
        \centering
        \includegraphics[trim={1.3cm 1.8cm 1.5cm 2.2cm},clip,width=0.49\textwidth]{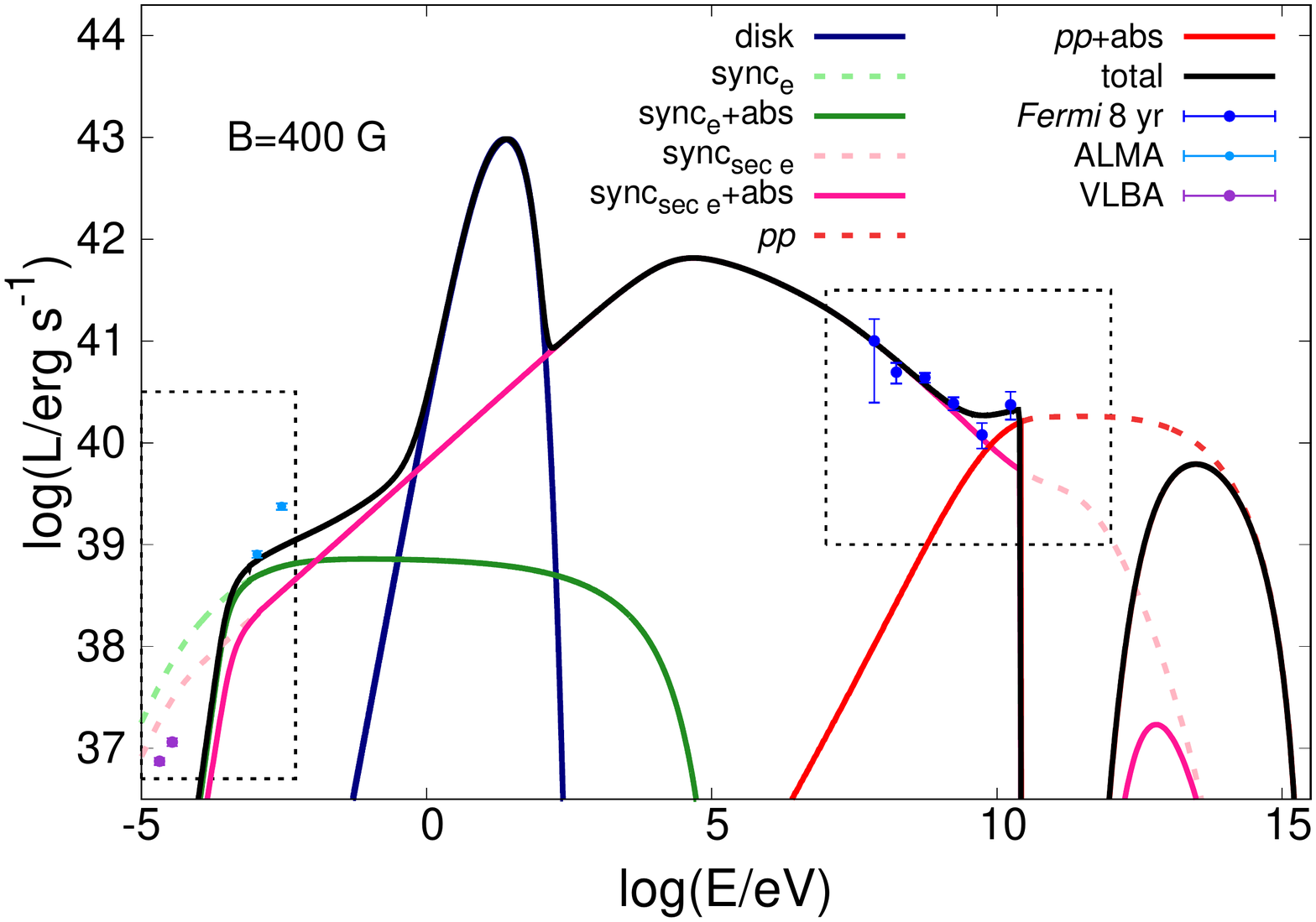}
        \caption{SED assuming $B=400$ G and $L_{p}/L_{e}=100$. The number of cloud impacts, $N_{\rm events}$, is determined by adjusting the model to match the observed total gamma luminosity. The gamma absorption is produced by the UV photons of the BLR and the optical photons from the accretion disk. The regions contained in the dashed rectangles are expanded in Fig. \ref{fig:NGC1068_radio} and Fig. \ref{fig:NGC1068_gamma}.}
        \label{fig:NGC1068_400}
\end{figure}

\begin{figure}
        \centering
        \includegraphics[trim={1.3cm 1.8cm 1.5cm 2.2cm},clip,width=0.49\textwidth]{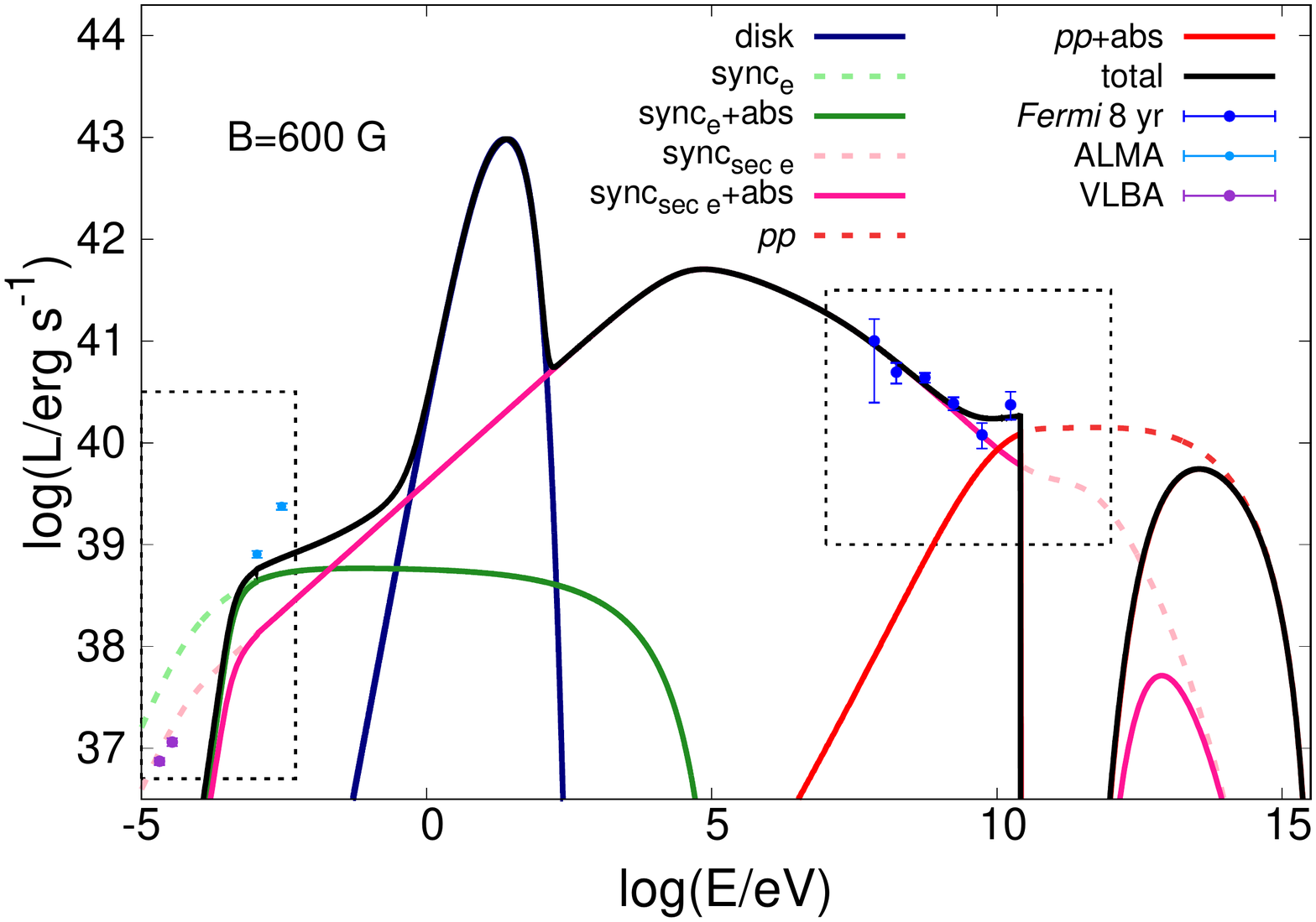}
        \caption{SED assuming $B=600$ G and $L_{p}/L_{e}=100$.  The number of cloud impacts, $N_{\rm events}$, is determined by adjusting the model to match the observed total gamma luminosity. The gamma absorption is produced by the UV photons of the BLR and the optical photons from the accretion disk. The regions contained in the dashed rectangles are expanded in Fig. \ref{fig:NGC1068_radio} and Fig. \ref{fig:NGC1068_gamma}.}
        \label{fig:NGC1068_600}
\end{figure}

Considering a magnetization of $10\%$ (Eq. \ref{eq:magnetization10}), we find in the case of $L_{p}/L_{e}=100$ that the required number of simultaneous $N_{\rm events}$ to match the gamma luminosity observed by \textit{Fermi} is $\sim 2.8\times10^{3}$. The luminosity in the range of the \textit{Swift} data is \mbox{$2.42\times10^{42}$ erg s$^{-1}$}, which is more than twice the emission measured, implying that the source should be $\sim 60\%$ obscured if the contribution of other sources in that band is negligible. The radio flux at $256$ GHz is overestimated by about $12\%$ (see Fig. \ref{fig:NGC1068_200}). In consequence, we calculate the SED for \mbox{$B=400$ G} and \mbox{$B=600$ G} to see whether higher magnetic fields improve the results. The corresponding magnetization ratios, maximum energies for the particles, and the luminosity in some bands are shown in Table \ref{tab:NGC1068models} for the two scenarios. The corresponding SEDs are presenting in Fig. \ref{fig:NGC1068_400} and Fig. \ref{fig:NGC1068_600}.

\begin{table}[htbp]
        \centering
        \caption{Parameters obtained with our model for $B=400$ G and $B=600$ G, constraining the number of events with the gamma luminosity.} 
        \begin{tabular}{l c c}
                \hline\hline 
                Parameter & \multicolumn{2}{c}{Magnetic field} \\
                          & 400 G & 600 G \\ \hline
                $\beta$ & $0.40$ & $0.90$ \\ 
                $E^{e}_{\rm max}$ & $2.6\times10^{10}$ eV & $2.1\times10^{10}$ eV \\
                $E^{p}_{\rm max}$ & $2.9\times10^{15}$ eV & $4.3\times10^{15}$ eV \\ \hline
                $N_{\rm events}$ & $1.9\times10^{3}$ & $1.5\times10^{3}$ \\
                $L_{256\,{\rm GHz}}$ & $6.90\times10^{38}$ erg s$^{-1}$ & $5.68\times10^{38}$ erg s$^{-1}$ \\
                $L_{694\,{\rm GHz}}$ & $9.55\times10^{38}$ erg s$^{-1}$ & $7.39\times10^{38}$ erg s$^{-1}$ \\
                $L_{14-195\,{\rm keV}}$ & $1.64\times10^{42}$ erg s$^{-1}$ & $1.25\times10^{42}$ erg s$^{-1}$ \\
                $L_{0.1-100\,{\rm GeV}}$ & $1.86\times10^{41}$ erg s$^{-1}$ & $1.79\times10^{41}$ erg s$^{-1}$ \\
 \hline
        \end{tabular}
        \label{tab:NGC1068models}
\end{table} 

\begin{figure}
        \centering
        \includegraphics[trim={0.3cm 0.7cm 2.7cm 0.3cm},clip,width=0.49\textwidth]{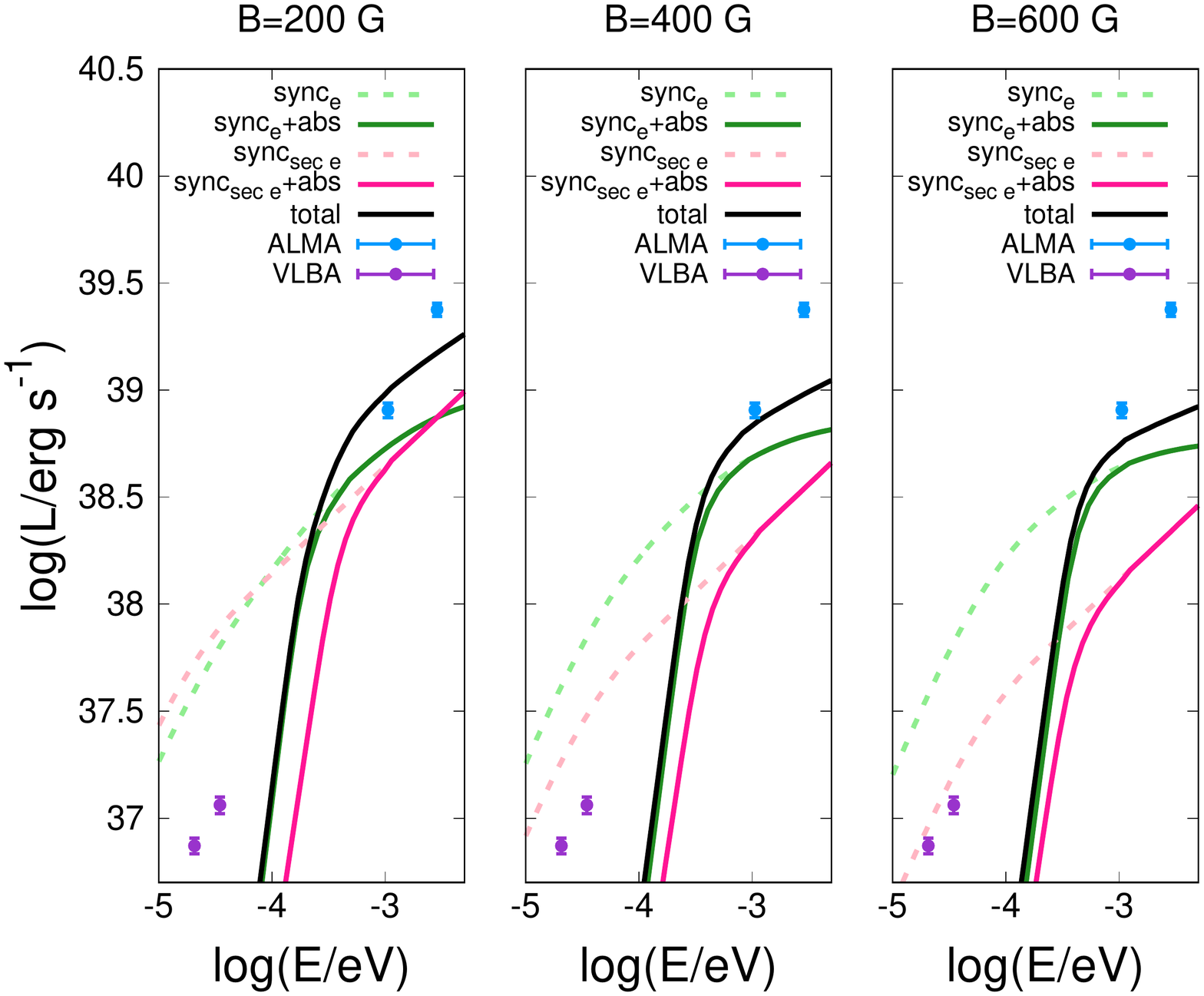}
        \caption{SEDs in the radio range for different values of the magnetic field assuming $L_{p}/L_{e}=100$. The number of cloud impacts, $N_{\rm events}$, is determined by adjusting the model to match the observed total gamma luminosity.}
        \label{fig:NGC1068_radio}
\end{figure}

\begin{figure}
        \centering
    \includegraphics[trim={0.3cm 0.7cm 2.7cm 0.3cm},clip,width=0.49\textwidth]{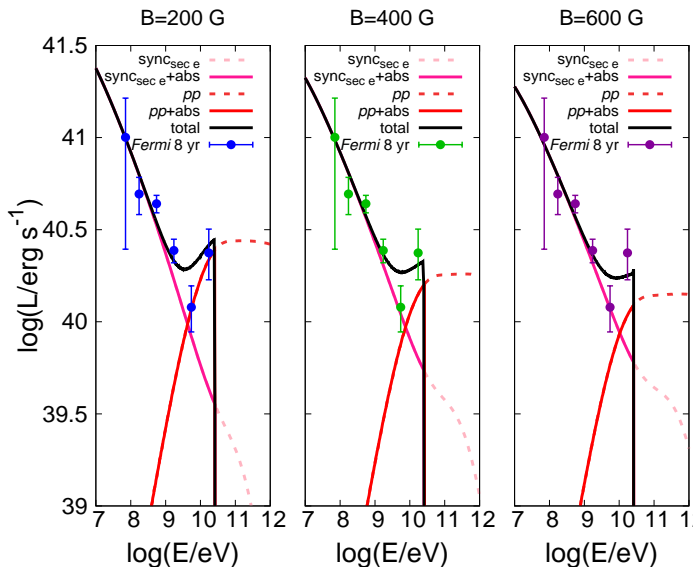}
        \caption{Fits of the SEDs in the gamma range for different values of the magnetic field assuming $L_{p}/L_{e}=100$. The number of cloud impacts, $N_{\rm events}$, is determined by adjusting the model to match the observed total gamma luminosity.}
        \label{fig:NGC1068_gamma}
\end{figure}

 We see  in all the cases that the VLBA limit is not exceeded because the radiation is strongly attenuated by SSA (see Fig. \ref{fig:NGC1068_radio}). With \mbox{$B=400$ G}, the total number of events required to reach the total gamma emission measured by \textit{Fermi} is $1.9\times10^{3}$, whereas with \mbox{$B=600$} G the presence of $1.5\times10^{3}$ simultaneous impacts is enough (see Table \ref{tab:NGC1068models} and Fig. \ref{fig:NGC1068_gamma}). The hard X-ray total luminosity predicted is \mbox{$1.64\times10^{42}$ erg s$^{-1}$} for \mbox{$B=400$ G}, and \mbox{$1.25\times10^{42}$ erg s$^{-1}$} for \mbox{$B=600$ G}. Therefore, the obscuration of the source should be at least between $20\%$ and $40\%$. These values can be increased by the existence of other sources emitting hard X-rays (e.g., a corona). Coronae have characteristic temperatures of $\sim10^{9}$ K and emit X-rays by Comptonization of photons from the accretion disk \citep[see][ and references therein]{Vieyro2012}. The expected luminosity of coronae can be similar or even up to a few of orders of magnitude higher than  produced in our scenario, in which case the obscuration percentage raises. The detection of the Fe K-alpha line in \object{NGC 1068} suggests the presence of a corona in this source, but the evidence is still  not conclusive \citep{Bauer2015,Marinucci2016,Inoue2019}. With these magnetic field values, the upper limits imposed by the observations with ALMA are not violated (see Fig. \ref{fig:NGC1068_radio} and Table \ref{tab:NGC1068models}).

\section{Discussion}\label{sec:discussion}

In Section \ref{sec:NGC1068}, we assume between $1.5\times10^{3}$ and $2.8\times10^{3}$ simultaneous events in order to achieve the emission observed by \textit{Fermi}. Many authors found that the number of clouds in the BLR should be $\sim 10^{8}$ or even larger \citep{Arav1997,Dietrich1999}. \citet{Abol2017} found that the total number could go up to $10^{18}$ depending on the value of the filling factor and the optical depth. 

Under the assumption that the clouds are uniformly distributed, their number per unit of volume ($n_{\rm clouds}$) will be just the total number of clouds ($N_{\rm clouds}$) divided by the volume of the BLR ($V_{\rm BLR}$). Calculating the characteristic radius as mentioned previously, for \object{NGC 1068} we find \mbox{$R_{\rm BLR}=3.36\times10^{16}$ cm}. Since the BLR can be thought of as a thin shell extending from $r$ to $R_{\rm BLR}$ (see Table \ref{tab:NGC1068parameters}), the resulting volume is \mbox{$V_{\rm BLR}\sim1.5\times10^{50}$ cm$^{3}$}. On the other hand, the number of impacts per unit of time is $\dot{N}_{\rm clouds}=n_{\rm clouds}\,v_{\rm c}\,\pi\,r^{2}$. Requiring that $N_{\rm events}=\dot{N}_{\rm clouds}\,t_{\rm ss}=1.5\times10^{3}$, $\dot{N}_{\rm clouds}$ becomes $0.06$, which means $N_{\rm clouds}$ should be $\sim 3\times10^{8}$. This value agrees very well with the number of clouds estimated from the observations. The characteristic luminosity fluctuation would be $\sqrt{N_{\rm events}}/N_{\rm events}\sim 2.6\%$ in $t_{\rm ss}/N_{\rm events}\sim 16$ s, assuming Poisson statistics \citep{delPalacio2019}. Increasing the number of events  to $2.8\times10^{3}$, the total number of clouds in the BLR should be $N_{\rm clouds}=6\times10^{8}$. The variability expected in this case is $\sqrt{N_{\rm events}}/N_{\rm events}\sim 1.8\%$ in $t_{\rm ss}/N_{\rm events}\sim 9$ s. Therefore, the emission produced by these events for any of the magnetic field values considered will be detected as continuous and our previous analysis becomes valid.

Long-term variations in the X-ray luminosity of radio-quiet AGNs, usually understood as changes in the size and properties of the corona, are not predicted by this model \citep[see, e.g.,][for a detailed discussion about X-ray variability in AGNs]{Soldi2014}. Nevertheless, the existence of a corona is not incompatible with the model here presented. Fluctuations in the observed X-ray emission from the impacts could be produced by variations in the absorbers in the line of sight. Strong modifications of the local environment, for instance due to a change in the accretion regimen, could also result in alterations of the X-ray luminosity, but the gamma emission should also be  affected.

Considering that in $pp$ collisions the neutrinos produced by charged pions take $\sim 5\%$ of the energy of the relativistic proton \citep{Lamastra2016}, it is possible to have neutrinos with energies in the detection range of IceCube. Therefore, and given the maximum energies achieved by the particles, the processes presented in this work might contribute to the spectrum reported by \citet{Icecube2019}.


\section{Summary and conclusions}

In this paper we investigated the acceleration of particles and the non-thermal emission produced by the collision of broad-line region clouds and the accretion disk in active galactic nuclei. We proposed as the acceleration mechanism the diffusive shock acceleration and calculated the maximum energies that the particles can reach. We found that, depending on the strength of the magnetic field, electrons can be accelerated up to $36$ GeV, whereas the proton maximum energy rises  to $\sim 4$ PeV. The energy losses for electrons are dominated by synchrotron, whereas $pp$ interactions dominate the cooling for the protons. The accelerated particles cool down locally and do not escape from the source.

We found that the emission of a single event cannot be detected as a flare, whereas the luminosity of multiple simultaneous events can explain the gamma radiation of \object{NGC 1068} if its nucleus is at least obscured between $20\%$ and $40\%$ at hard X-ray frequencies. The high-energy gamma photons produced in $pp$ inelastic collisions are absorbed in the BLR radiation field, injecting secondary electrons. These secondaries emit synchrotron radiation in the detection range of \textit{Fermi}.

The number of simultaneous events needed to account for the gamma rays observed varies between $1.5\times10^{3}$ and $2.8\times10^{3}$, depending on the magnetic field assumed. These numbers are feasible if the total number of BLR clouds is between $3\times10^{8}$ and $6\times10^{8}$. The variability of luminosity in time generated because of the superposition of sources is too small to be detected. Given the maximum energies achieved by the protons, neutrinos with energies in the detection range of IceCube might be created in the collision of BLR clouds with accretion disks.

All in all, the model presented is an attractive alternative to explain the high-energy emission in active systems deprived of powerful jets. Further observations with the next generation of X-ray and gamma satellites \citep[e.g., \textit{ATHENA}, the sucessor of \textit{e-ASTROGAM};][]{Barcons2017,Rando2019} might contribute to validating and distinguishing our model from other possible scenarios \citep[e.g.,][]{Lamastra2019,Inoue2019}.

\begin{acknowledgements}

 We would like to thank the anonymous reviewer for suggestions and comments. ALM thanks to S. del Palacio for fruitful discussions, and to A. Streich for his technical support. This work was supported by the Argentine agencies CONICET (PIP 2014-00338), ANPCyT (PICT 2017-2865) and the Spanish Ministerio de Econom\'{i}a y Competitividad (MINECO/FEDER, UE) under grant  AYA2016-76012-C3-1-P.

\end{acknowledgements}

%
%

\bibliographystyle{aa} 
\bibliography{BLRC-AA} 

\begin{thebibliography}{82}
\expandafter\ifx\csname natexlab\endcsname\relax\def\natexlab#1{#1}\fi

\bibitem[{{Abolmasov} \& {Poutanen}(2017)}]{Abol2017}
{Abolmasov}, P. \& {Poutanen}, J. 2017, \mnras, 464, 152

\bibitem[{{Ackermann} {et~al.}(2012){Ackermann}, {Ajello}, {Allafort},
  {Baldini}, {Ballet}, {Bastieri}, {Bechtol}, {Bellazzini}, {Berenji}, {Bloom},
  {Bonamente}, {Borgland }, {Bouvier}, {Bregeon}, {Brigida}, {Bruel},
  {Buehler}, {Buson}, {Caliandro}, {Cameron}, {Caraveo}, {Casandjian},
  {Cecchi}, {Charles}, {Chekhtman}, {Cheung}, {Chiang}, {Cillis}, {Ciprini},
  {Claus}, {Cohen-Tanugi}, {Conrad}, {Cutini}, {de Palma}, {Dermer}, {Digel},
  {Silva}, {Drell}, {Drlica-Wagner}, {Favuzzi}, {Fegan}, {Fortin}, {Fukazawa},
  {Funk}, {Fusco}, {Gargano}, {Gasparrini}, {Germani}, {Giglietto}, {Giordano},
  {Glanzman}, {Godfrey}, {Grenier}, {Guiriec}, {Gustafsson}, {Hadasch},
  {Hayashida}, {Hays}, {Hughes}, {J{\'o}hannesson}, {Johnson}, {Kamae},
  {Katagiri}, {Kataoka}, {Kn{\"o}dlseder}, {Kuss}, {Lande}, {Longo}, {Loparco},
  {Lott}, {Lovellette}, {Lubrano}, {Madejski}, {Martin}, {Mazziotta},
  {McEnery}, {Michelson}, {Mizuno}, {Monte}, {Monzani}, {Morselli},
  {Moskalenko}, {Murgia}, {Nishino}, {Norris}, {Nuss}, {Ohno}, {Ohsugi},
  {Okumura}, {Omodei}, {Orlando}, {Ozaki}, {Parent}, {Persic}, {Pesce-Rollins},
  {Petrosian}, {Pierbattista}, {Piron}, {Pivato}, {Porter}, {Rain{\`o}},
  {Rando}, {Razzano}, {Reimer}, {Reimer}, {Ritz}, {Roth}, {Sbarra}, {Sgr{\`o}},
  {Siskind}, {Spandre}, {Spinelli}, {Stawarz}, {Strong}, {Takahashi}, {Tanaka},
  {Thayer}, {Tibaldo}, {Tinivella}, {Torres}, {Tosti}, {Troja}, {Uchiyama},
  {Vandenbroucke}, {Vianello}, {Vitale}, {Waite}, {Wood}, \&
  {Yang}}]{Ackermann2012}
{Ackermann}, M., {Ajello}, M., {Allafort}, A., {et~al.} 2012, \apj, 755, 164

\bibitem[{{Aharonian} \& {Atoyan}(2000)}]{Aharonian2000}
{Aharonian}, F.~A. \& {Atoyan}, A.~M. 2000, \aap, 362, 937

\bibitem[{{Antonucci}(1993)}]{Antonucci1993}
{Antonucci}, R. 1993, \araa, 31, 473

\bibitem[{{Antonucci}(1984)}]{Antonucci1984}
{Antonucci}, R.~R.~J. 1984, \apj, 278, 499

\bibitem[{{Antonucci} \& {Miller}(1985)}]{AntonucciMiller1985}
{Antonucci}, R.~R.~J. \& {Miller}, J.~S. 1985, \apj, 297, 621

\bibitem[{{Araudo} {et~al.}(2010){Araudo}, {Bosch-Ramon}, \&
  {Romero}}]{Araudo2010}
{Araudo}, A.~T., {Bosch-Ramon}, V., \& {Romero}, G.~E. 2010, \aap, 522, A97

\bibitem[{{Arav} {et~al.}(1997){Arav}, {Barlow}, {Laor}, \& {Bland
  ford}}]{Arav1997}
{Arav}, N., {Barlow}, T.~A., {Laor}, A., \& {Bland ford}, R.~D. 1997, \mnras,
  288, 1015

\bibitem[{{Armitage} {et~al.}(1996){Armitage}, {Zurek}, \&
  {Davies}}]{Armitage1996}
{Armitage}, P.~J., {Zurek}, W.~H., \& {Davies}, M.~B. 1996, \apj, 470, 237

\bibitem[{{Barcons} {et~al.}(2017){Barcons}, {Barret}, {Decourchelle}, {den
  Herder}, {Fabian}, {Matsumoto}, {Lumb}, {Nandra}, {Piro}, {Smith}, \&
  {Willingale}}]{Barcons2017}
{Barcons}, X., {Barret}, D., {Decourchelle}, A., {et~al.} 2017, Astronomische
  Nachrichten, 338, 153

\bibitem[{{Bauer} {et~al.}(2015){Bauer}, {Ar{\'e}valo}, {Walton}, {Koss},
  {Puccetti}, {Gandhi}, {Stern}, {Alexander}, {Balokovi{\'c}}, {Boggs},
  {Brandt}, {Brightman}, {Christensen}, {Comastri}, {Craig}, {Del Moro},
  {Hailey}, {Harrison}, {Hickox}, {Luo}, {Markwardt}, {Marinucci}, {Matt},
  {Rigby}, {Rivers}, {Saez}, {Treister}, {Urry}, \& {Zhang}}]{Bauer2015}
{Bauer}, F.~E., {Ar{\'e}valo}, P., {Walton}, D.~J., {et~al.} 2015, \apj, 812,
  116

\bibitem[{{Bell}(1978)}]{Bell1978}
{Bell}, A.~R. 1978, MNRAS, 182, 147\textendash 156

\bibitem[{{Blandford} {et~al.}(1990){Blandford}, {Netzer}, {Woltjer},
  {Courvoisier}, \& {Mayor}}]{Blandford1990}
{Blandford}, R.~D., {Netzer}, H., {Woltjer}, L., {Courvoisier}, T.~J.-L., \&
  {Mayor}, M., eds. 1990, {Active Galactic Nuclei}, 97

\bibitem[{{Blandford} \& {Ostriker}(1978)}]{Blandford1978}
{Blandford}, R.~D. \& {Ostriker}, J.~P. 1978, ApJL, 221, L29\textendash L32

\bibitem[{{Blumenthal} \& {Gould}(1970)}]{Blumenthal1970}
{Blumenthal}, G.~R. \& {Gould}, R.~J. 1970, Rev. Mod. Phy., 42, 237\textendash
  271

\bibitem[{{Bosch-Ramon} {et~al.}(2010){Bosch-Ramon}, {Romero}, {Araudo}, \&
  {Paredes}}]{Bosch-Ramon2010}
{Bosch-Ramon}, V., {Romero}, G.~E., {Araudo}, A.~T., \& {Paredes}, J.~M. 2010,
  \aap, 511, A8

\bibitem[{{B{\"o}ttcher} \& {Els}(2016)}]{Boettcher2016}
{B{\"o}ttcher}, M. \& {Els}, P. 2016, \apj, 821, 102

\bibitem[{{Cox}(2000)}]{Allen}
{Cox}, A.~N. 2000, Allen's astrophysical quantities

\bibitem[{{del Palacio} {et~al.}(2019){del Palacio}, {Bosch-Ramon}, \&
  {Romero}}]{delPalacio2019}
{del Palacio}, S., {Bosch-Ramon}, V., \& {Romero}, G.~E. 2019, \aap, 623, A101

\bibitem[{{del Valle} {et~al.}(2018){del Valle}, {M{\"u}ller}, \&
  {Romero}}]{delValle2018}
{del Valle}, M.~V., {M{\"u}ller}, A.~L., \& {Romero}, G.~E. 2018, \mnras, 475,
  4298

\bibitem[{{Dermer} \& {Giebels}(2016)}]{Dermer2016}
{Dermer}, C.~D. \& {Giebels}, B. 2016, Comptes Rendus Physique, 17, 594

\bibitem[{{Dietrich} {et~al.}(1999){Dietrich}, {Wagner}, {Courvoisier}, {Bock},
  \& {North}}]{Dietrich1999}
{Dietrich}, M., {Wagner}, S.~J., {Courvoisier}, T.~J.~L., {Bock}, H., \&
  {North}, P. 1999, \aap, 351, 31

\bibitem[{{D{\"o}nmez}(2006)}]{Doenmez2006}
{D{\"o}nmez}, O. 2006, \apss, 305, 187

\bibitem[{{Doroshenko} {et~al.}(2012){Doroshenko}, {Sergeev}, {Klimanov},
  {Pronik}, \& {Efimov}}]{Doroshenko2012}
{Doroshenko}, V.~T., {Sergeev}, S.~G., {Klimanov}, S.~A., {Pronik}, V.~I., \&
  {Efimov}, Y.~S. 2012, \mnras, 426, 416

\bibitem[{{Drury}(1983)}]{Drury1983}
{Drury}, L.~O. 1983, Rep. Prog. Phys., 46, 973\textendash1027

\bibitem[{{Fabian}(1999)}]{Fabian1999}
{Fabian}, A.~C. 1999, Proceedings of the National Academy of Science, 96, 4749

\bibitem[{{Frank} {et~al.}(2002){Frank}, {King}, \& {Raine}}]{APAp}
{Frank}, J., {King}, A., \& {Raine}, D.~J. 2002, Accretion Power in
  Astrophysics: Third Edition

\bibitem[{{Gallimore} {et~al.}(2004){Gallimore}, {Baum}, \&
  {O'Dea}}]{Gallimore2004}
{Gallimore}, J.~F., {Baum}, S.~A., \& {O'Dea}, C.~P. 2004, \apj, 613, 794

\bibitem[{{Garc{\'\i}a-Burillo} {et~al.}(2016){Garc{\'\i}a-Burillo}, {Combes},
  {Ramos Almeida}, {Usero}, {Krips}, {Alonso-Herrero}, {Aalto}, {Casasola},
  {Hunt}, {Mart{\'\i}n}, {Viti}, {Colina}, {Costagliola}, {Eckart}, {Fuente},
  {Henkel}, {M{\'a}rquez}, {Neri}, {Schinnerer}, {Tacconi}, \& {van der
  Werf}}]{Garcia2016}
{Garc{\'\i}a-Burillo}, S., {Combes}, F., {Ramos Almeida}, C., {et~al.} 2016,
  \apjl, 823, L12

\bibitem[{{Ginzburg} \& {Syrovatskii}(1964)}]{Ginzburg1964}
{Ginzburg}, V.~L. \& {Syrovatskii}, S.~I. 1964, The Origin of Cosmic Rays
  (Macmillan)

\bibitem[{{Gould} \& {Schr{\'e}der}(1967)}]{Gould1967}
{Gould}, R.~J. \& {Schr{\'e}der}, G.~P. 1967, Physical Review, 155, 1404

\bibitem[{{Grier} {et~al.}(2013){Grier}, {Peterson}, {Horne}, {Bentz}, {Pogge},
  {Denney}, {De Rosa}, {Martini}, {Kochanek}, {Zu}, {Shappee}, {Siverd},
  {Beatty}, {Sergeev}, {Kaspi}, {Araya Salvo}, {Bird}, {Bord}, {Borman}, {Che},
  {Chen}, {Cohen}, {Dietrich}, {Doroshenko}, {Efimov}, {Free}, {Ginsburg},
  {Henderson}, {King}, {Mogren}, {Molina}, {Mosquera}, {Nazarov}, {Okhmat},
  {Pejcha}, {Rafter}, {Shields}, {Skowron}, {Szczygiel}, {Valluri}, \& {van
  Saders}}]{Grier2013}
{Grier}, C.~J., {Peterson}, B.~M., {Horne}, K., {et~al.} 2013, \apj, 764, 47

\bibitem[{{Hillas}(1984)}]{Hillas1984}
{Hillas}, A.~M. 1984, ARA\&A, 22, 425\textendash444

\bibitem[{{IceCube Collaboration} {et~al.}(2019){IceCube Collaboration},
  {Aartsen}, {Ackermann}, {Adams}, {Aguilar}, {Ahlers}, {Ahrens}, {Alispach},
  {Andeen}, {Anderson}, {Ansseau}, {Anton}, {Arg{\"u}elles}, {Auffenberg},
  {Axani}, {Backes}, {Bagherpour}, {Bai}, {Balagopal V.}, {Barbano}, {Barwick},
  {Bastian}, {Baum}, {Baur}, {Bay}, {Beatty}, {Becker}, {Becker Tjus},
  {BenZvi}, {Berley}, {Bernardini}, {Besson}, {Binder}, {Bindig}, {Blaufuss},
  {Blot}, {Bohm}, {B{\"o}rner}, {B{\"o}ser}, {Botner}, {B{\"o}ttcher},
  {Bourbeau}, {Bourbeau}, {Bradascio}, {Braun}, {Bron}, {Brostean-Kaiser},
  {Burgman}, {Buscher}, {Busse}, {Carver}, {Chen}, {Cheung}, {Chirkin}, {Choi},
  {Clark}, {Classen}, {Coleman}, {Collin}, {Conrad}, {Coppin}, {Correa},
  {Cowen}, {Cross}, {Dave}, {De Clercq}, {DeLaunay}, {Dembinski}, {Deoskar},
  {De Ridder}, {Desiati}, {de Vries}, {de Wasseige}, {de With}, {DeYoung},
  {Diaz}, {D{\'\i}az-V{\'e}lez}, {Dujmovic}, {Dunkman}, {Dvorak}, {Eberhardt},
  {Ehrhardt}, {Eller}, {Engel}, {Evenson}, {Fahey}, {Fazely}, {Felde},
  {Filimonov}, {Finley}, {Fox}, {Franckowiak}, {Friedman}, {Fritz}, {Gaisser},
  {Gallagher}, {Ganster}, {Garrappa}, {Gerhardt}, {Ghorbani}, {Glauch},
  {Gl{\"u}senkamp}, {Goldschmidt}, {Gonzalez}, {Grant}, {Griffith}, {Griswold},
  {G{\"u}nder}, {G{\"u}nd{\"u}z}, {Haack}, {Hallgren}, {Halliday}, {Halve},
  {Halzen}, {Hanson}, {Haungs}, {Hebecker}, {Heereman}, {Heix}, {Helbing},
  {Hellauer}, {Henningsen}, {Hickford}, {Hignight}, {Hill}, {Hoffman},
  {Hoffmann}, {Hoinka}, {Hokanson-Fasig}, {Hoshina}, {Huang}, {Huber}, {Huber},
  {Hultqvist}, {H{\"u}nnefeld}, {Hussain}, {In}, {Iovine}, {Ishihara},
  {Japaridze}, {Jeong}, {Jero}, {Jones}, {Jonske}, {Joppe}, {Kang}, {Kang},
  {Kappes}, {Kappesser}, {Karg}, {Karl}, {Karle}, {Katz}, {Kauer}, {Kelley},
  {Kheirandish}, {Kim}, {Kintscher}, {Kiryluk}, {Kittler}, {Klein}, {Koirala},
  {Kolanoski}, {K{\"o}pke}, {Kopper}, {Kopper}, {Koskinen}, {Kowalski},
  {Krings}, {Kr{\"u}ckl}, {Kulacz}, {Kurahashi}, {Kyriacou}, {Lanfranchi},
  {Larson}, {Lauber}, {Lazar}, {Leonard}, {Leszczy{\'n}ska}, {Leuermann},
  {Liu}, {Lohfink}, {Lozano Mariscal}, {Lu}, {Lucarelli}, {L{\"u}nemann},
  {Luszczak}, {Lyu}, {Ma}, {Madsen}, {Maggi}, {Mahn}, {Makino}, {Mallik},
  {Mallot}, {Mancina}, {Mari\{{\textcommabelow s}\}}, {Maruyama}, {Mase},
  {Maunu}, {McNally}, {Meagher}, {Medici}, {Medina}, {Meier}, {Meighen-Berger},
  {Menne}, {Merino}, {Meures}, {Micallef}, {Mockler}, {Moment{\'e}},
  {Montaruli}, {Moore}, {Morse}, {Moulai}, {Muth}, {Nagai}, {Naumann}, {Neer},
  {Niederhausen}, {Nisa}, {Nowicki}, {Nygren}, {Obertacke Pollmann}, {Oehler},
  {Olivas}, {O'Murchadha}, {O'Sullivan}, {Palczewski}, {Pandya}, {Pankova},
  {Park}, {Peiffer}, {P{\'e}rez de los Heros}, {Philippen}, {Pieloth}, {Pinat},
  {Pizzuto}, {Plum}, {Porcelli}, {Price}, {Przybylski}, {Raab}, {Raissi},
  {Rameez}, {Rauch}, {Rawlins}, {Rea}, {Reimann}, {Relethford}, {Renschler},
  {Renzi}, {Resconi}, {Rhode}, {Richman}, {Robertson}, {Rongen}, {Rott},
  {Ruhe}, {Ryckbosch}, {Rysewyk}, {Safa}, {Sanchez Herrera}, {Sandrock},
  {Sandroos}, {Santander}, {Sarkar}, {Sarkar}, {Satalecka}, {Schaufel},
  {Schieler}, {Schlunder}, {Schmidt}, {Schneider}, {Schneider}, {Schr{\"o}der},
  {Schumacher}, {Sclafani}, {Seckel}, {Seunarine}, {Shefali}, {Silva},
  {Snihur}, {Soedingrekso}, {Soldin}, {Song}, {Spiczak}, {Spiering},
  {Stachurska}, {Stamatikos}, {Stanev}, {Stein}, {Steinm{\"u}ller}, {Stettner},
  {Steuer}, {Stezelberger}, {Stokstad}, {St{\"o}{\ss}l}, {Strotjohann},
  {St{\"u}rwald}, {Stuttard}, {Sullivan}, {Taboada}, {Tenholt}, {Ter-Antonyan},
  {Terliuk}, {Tilav}, {Tollefson}, {Tomankova}, {T{\"o}nnis}, {Toscano},
  {Tosi}, {Trettin}, {Tselengidou}, {Tung}, {Turcati}, {Turcotte}, {Turley},
  {Ty}, {Unger}, {Unland Elorrieta}, {Usner}, {Vandenbroucke}, {Van Driessche},
  {van Eijk}, {van Eijndhoven}, {van Santen}, {Verpoest}, {Vraeghe}, {Walck},
  {Wallace}, {Wallraff}, {Wandkowsky}, {Watson}, {Weaver}, {Weindl}, {Weiss},
  {Weldert}, {Wendt}, {Werthebach}, {Whelan}, {Whitehorn}, {Wiebe}, {Wiebusch},
  {Wille}, {Williams}, {Wills}, {Wolf}, {Wood}, {Wood}, {Woschnagg}, {Wrede},
  {Xu}, {Xu}, {Xu}, {Yanez}, {Yodh}, {Yoshida}, {Yuan}, \&
  {Z{\"o}cklein}}]{Icecube2019}
{IceCube Collaboration}, {Aartsen}, M.~G., {Ackermann}, M., {et~al.} 2019,
  arXiv e-prints, arXiv:1910.08488

\bibitem[{{Impellizzeri} {et~al.}(2019){Impellizzeri}, {Gallimore}, {Baum},
  {Elitzur}, {Davies}, {Lutz}, {Maiolino}, {Marconi}, {Nikutta}, {O'Dea}, \&
  {Sani}}]{Impellizzeri2019}
{Impellizzeri}, C.~M.~V., {Gallimore}, J.~F., {Baum}, S.~A., {et~al.} 2019,
  \apjl, 884, L28

\bibitem[{{Inoue} {et~al.}(2019){Inoue}, {Khangulyan}, \& {Doi}}]{Inoue2019}
{Inoue}, Y., {Khangulyan}, D., \& {Doi}, A. 2019, arXiv e-prints,
  arXiv:1909.02239

\bibitem[{{Kaspi} {et~al.}(2007){Kaspi}, {Brandt}, {Maoz}, {Netzer},
  {Schneider}, \& {Shemmer}}]{Kaspi2007}
{Kaspi}, S., {Brandt}, W.~N., {Maoz}, D., {et~al.} 2007, \apj, 659, 997

\bibitem[{{Kelner} {et~al.}(2006){Kelner}, {Aharonian}, \&
  {Bugayov}}]{Kelner2006}
{Kelner}, S.~R., {Aharonian}, F.~A., \& {Bugayov}, V.~V. 2006, Phys. Rev. D, 74

\bibitem[{{Komissarov} \& {Barkov}(2007)}]{Komissarov2007}
{Komissarov}, S.~S. \& {Barkov}, M.~V. 2007, \mnras, 382, 1029

\bibitem[{{Lamastra} {et~al.}(2016){Lamastra}, {Fiore}, {Guetta}, {Antonelli},
  {Colafrancesco}, {Menci}, {Puccetti}, {Stamerra}, \&
  {Zappacosta}}]{Lamastra2016}
{Lamastra}, A., {Fiore}, F., {Guetta}, D., {et~al.} 2016, \aap, 596, A68

\bibitem[{{Lamastra} {et~al.}(2019){Lamastra}, {Tavecchio}, {Romano},
  {Landoni}, \& {Vercellone}}]{Lamastra2019}
{Lamastra}, A., {Tavecchio}, F., {Romano}, P., {Landoni}, M., \& {Vercellone},
  S. 2019, Astroparticle Physics, 112, 16

\bibitem[{Landau \& Lifshitz(1959)}]{LL1959}
Landau, L.~D. \& Lifshitz, E. 1959, Fluid Mechanics

\bibitem[{{Laor}(2003)}]{Laor2003}
{Laor}, A. 2003, \apj, 590, 86

\bibitem[{{Lee} {et~al.}(1996){Lee}, {Kang}, \& {Ryu}}]{LK1996}
{Lee}, H.~M., {Kang}, H., \& {Ryu}, D. 1996, \apj, 464, 131

\bibitem[{{Lodato} \& {Bertin}(2003)}]{LodatoBertin2003}
{Lodato}, G. \& {Bertin}, G. 2003, \aap, 398, 517

\bibitem[{{Marinucci} {et~al.}(2016){Marinucci}, {Bianchi}, {Matt},
  {Alexander}, {Balokovi{\'c}}, {Bauer}, {Brandt}, {Gand hi}, {Guainazzi},
  {Harrison}, {Iwasawa}, {Koss}, {Madsen}, {Nicastro}, {Puccetti}, {Ricci},
  {Stern}, \& {Walton}}]{Marinucci2016}
{Marinucci}, A., {Bianchi}, S., {Matt}, G., {et~al.} 2016, \mnras, 456, L94

\bibitem[{{Marinucci} {et~al.}(2012){Marinucci}, {Bianchi}, {Nicastro}, {Matt},
  \& {Goulding}}]{Marinucci2012}
{Marinucci}, A., {Bianchi}, S., {Nicastro}, F., {Matt}, G., \& {Goulding},
  A.~D. 2012, \apj, 748, 130

\bibitem[{{Myasnikov} {et~al.}(1998){Myasnikov}, {Zhekov}, \&
  {Belov}}]{MZB1998}
{Myasnikov}, A.~V., {Zhekov}, S.~A., \& {Belov}, N.~A. 1998, MNRAS, 298, 1021
  \textendash 1029

\bibitem[{{Nayakshin} {et~al.}(2004){Nayakshin}, {Cuadra}, \&
  {Sunyaev}}]{Nayakshin2004}
{Nayakshin}, S., {Cuadra}, J., \& {Sunyaev}, R. 2004, \aap, 413, 173

\bibitem[{{O'C Drury} {et~al.}(1996){O'C Drury}, {Duffy}, \&
  {Kirk}}]{Drury1996}
{O'C Drury}, L., {Duffy}, P., \& {Kirk}, J.~G. 1996, \aap, 309, 1002

\bibitem[{{Padmanabhan}(2002)}]{PT2002}
{Padmanabhan}, T. 2002, Theoretical Astrophysics - Volume 3, Galaxies and
  Cosmology, 638

\bibitem[{{Peterson}(1998)}]{Peterson1998}
{Peterson}, B.~M. 1998, Advances in Space Research, 21, 57

\bibitem[{{Peterson} \& {Wandel}(1999)}]{PetersonW1999}
{Peterson}, B.~M. \& {Wandel}, A. 1999, \apjl, 521, L95

\bibitem[{{Pier} {et~al.}(1994){Pier}, {Antonucci}, {Hurt}, {Kriss}, \&
  {Krolik}}]{Pier1994}
{Pier}, E.~A., {Antonucci}, R., {Hurt}, T., {Kriss}, G., \& {Krolik}, J. 1994,
  \apj, 428, 124

\bibitem[{{Ramos Almeida} {et~al.}(2016){Ramos Almeida}, {Mart{\'\i}nez
  Gonz{\'a}lez}, {Asensio Ramos}, {Acosta-Pulido}, {H{\"o}nig},
  {Alonso-Herrero}, {Tadhunter}, \&
  {Gonz{\'a}lez-Mart{\'\i}n}}]{RamosAlmeida2016}
{Ramos Almeida}, C., {Mart{\'\i}nez Gonz{\'a}lez}, M.~J., {Asensio Ramos}, A.,
  {et~al.} 2016, \mnras, 461, 1387

\bibitem[{{Rando} {et~al.}(2019){Rando}, {De Angelis}, {Mallamaci}, \&
  {e-ASTROGAM collaboration}}]{Rando2019}
{Rando}, R., {De Angelis}, A., {Mallamaci}, M., \& {e-ASTROGAM collaboration}.
  2019, in Journal of Physics Conference Series, Vol. 1181, Journal of Physics
  Conference Series, 012044

\bibitem[{Raymond {et~al.}(1976)Raymond, Cox, \& Smith}]{RCS1976}
Raymond, J., Cox, P.~D., \& Smith, B.~W. 1976, ApJ, 204, 290

\bibitem[{{Romero} {et~al.}(2010{\natexlab{a}}){Romero}, {Del Valle}, \&
  {Orellana}}]{Romero2010b}
{Romero}, G.~E., {Del Valle}, M.~V., \& {Orellana}, M. 2010{\natexlab{a}},
  \aap, 518, A12

\bibitem[{Romero \& Paredes(2011)}]{Romero2011}
Romero, G.~E. \& Paredes, J.~M. 2011, Introducci\'on a la Astrof\'isica
  Relativista (Publicacions i Edicions de la Universitat de Barcelona)

\bibitem[{{Romero} {et~al.}(2010{\natexlab{b}}){Romero}, {Vieyro}, \&
  {Vila}}]{Romero2010}
{Romero}, G.~E., {Vieyro}, F.~L., \& {Vila}, G.~S. 2010{\natexlab{b}}, \aap,
  519, A109

\bibitem[{{Rybicki} \& {Lightman}(1985)}]{Rybicki}
{Rybicki}, G.~B. \& {Lightman}, A.~P. 1985, {Radiative processes in
  astrophysics.}

\bibitem[{{Santillan} {et~al.}(2004){Santillan}, {Franco}, \&
  {Kim}}]{Santillan2004}
{Santillan}, A., {Franco}, J., \& {Kim}, J. 2004, Journal of Korean
  Astronomical Society, 37, 233

\bibitem[{{Schlickeiser}(2002)}]{Schlickeiser2002}
{Schlickeiser}, R. 2002, {Cosmic Ray Astrophysics}

\bibitem[{{Shadmehri}(2015)}]{S2015}
{Shadmehri}, M. 2015, MNRAS, 451, 3671 \textendash 3678

\bibitem[{Shakura \& Sunyaev(1973)}]{SS1973}
Shakura, N.~I. \& Sunyaev, R.~A. 1973, \aap, 24, 337

\bibitem[{{Shin} {et~al.}(2008){Shin}, {Stone}, \& {Snyder}}]{Shin2008}
{Shin}, M.-S., {Stone}, J.~M., \& {Snyder}, G.~F. 2008, \apj, 680, 336

\bibitem[{{Sillanpaa} {et~al.}(1988){Sillanpaa}, {Haarala}, {Valtonen},
  {Sundelius}, \& {Byrd}}]{Sillanpaa1988}
{Sillanpaa}, A., {Haarala}, S., {Valtonen}, M.~J., {Sundelius}, B., \& {Byrd},
  G.~G. 1988, \apj, 325, 628

\bibitem[{{Soldi} {et~al.}(2014){Soldi}, {Beckmann}, {Baumgartner}, {Ponti},
  {Shrader}, {Lubi{\'n}ski}, {Krimm}, {Mattana}, \& {Tueller}}]{Soldi2014}
{Soldi}, S., {Beckmann}, V., {Baumgartner}, W.~H., {et~al.} 2014, \aap, 563,
  A57

\bibitem[{{Syer} {et~al.}(1991){Syer}, {Clarke}, \& {Rees}}]{Syer1991}
{Syer}, D., {Clarke}, C.~J., \& {Rees}, M.~J. 1991, \mnras, 250, 505

\bibitem[{Tenorio-Tagle(1980)}]{TT1980}
Tenorio-Tagle, G. 1980, A\&A, 94, 338\textendash344

\bibitem[{{The Fermi-LAT collaboration}(2019)}]{Fermi2019}
{The Fermi-LAT collaboration}. 2019, arXiv e-prints, arXiv:1902.10045

\bibitem[{{Treves} {et~al.}(1988){Treves}, {Maraschi}, \&
  {Abramowicz}}]{TMA1988}
{Treves}, A., {Maraschi}, L., \& {Abramowicz}, M. 1988, \pasp, 100, 427

\bibitem[{{Tully}(1988)}]{Tully1988}
{Tully}, R.~B. 1988, {Nearby galaxies catalog}

\bibitem[{{Urry} \& {Padovani}(1995)}]{Urry1995}
{Urry}, C.~M. \& {Padovani}, P. 1995, \pasp, 107, 803

\bibitem[{{Valtonen} {et~al.}(2008){Valtonen}, {Lehto}, {Nilsson}, {Heidt},
  {Takalo}, {Sillanp{\"a}{\"a}}, {Villforth}, {Kidger}, {Poyner}, {Pursimo},
  {Zola}, {Wu}, {Zhou}, {Sadakane}, {Drozdz}, {Koziel}, {Marchev}, {Ogloza},
  {Porowski}, {Siwak}, {Stachowski}, {Winiarski}, {Hentunen}, {Nissinen},
  {Liakos}, \& {Dogru}}]{Valtonen2008}
{Valtonen}, M.~J., {Lehto}, H.~J., {Nilsson}, K., {et~al.} 2008, \nat, 452, 851

\bibitem[{{Vieyro} \& {Romero}(2012)}]{Vieyro2012}
{Vieyro}, F.~L. \& {Romero}, G.~E. 2012, \aap, 542, A7

\bibitem[{{Vink} \& {Yamazaki}(2014)}]{Vink2014}
{Vink}, J. \& {Yamazaki}, R. 2014, \apj, 780, 125

\bibitem[{{Wojaczy{\'n}ski} {et~al.}(2015){Wojaczy{\'n}ski}, {Nied{\'z}wiecki},
  {Xie}, \& {Szanecki}}]{Woja2015}
{Wojaczy{\'n}ski}, R., {Nied{\'z}wiecki}, A., {Xie}, F.-G., \& {Szanecki}, M.
  2015, \aap, 584, A20

\bibitem[{{Xie} {et~al.}(2009){Xie}, {Ma}, {Zhang}, {Du}, {Hao}, {Yi}, \&
  {Qiao}}]{Xie2009}
{Xie}, Z.~H., {Ma}, L., {Zhang}, X., {et~al.} 2009, \apj, 707, 866

\bibitem[{{Zaino} {et~al.}(2020){Zaino}, {Bianchi}, {Marinucci}, {Matt},
  {Bauer}, {Brandt}, {Gandhi}, {Guainazzi}, {Iwasawa}, {Puccetti}, {Ricci}, \&
  {Walton}}]{Zaino2020}
{Zaino}, A., {Bianchi}, S., {Marinucci}, A., {et~al.} 2020, \mnras, 104

\bibitem[{{Zentsova}(1983)}]{Zentsova1983}
{Zentsova}, A.~S. 1983, \apss, 95, 11

\bibitem[{{Zurek} {et~al.}(1994){Zurek}, {Siemiginowska}, \&
  {Colgate}}]{Colgate1994}
{Zurek}, W.~H., {Siemiginowska}, A., \& {Colgate}, S.~A. 1994, \apj, 434, 46

\end{thebibliography}

\end{document}